# Stochastic AC Network-constrained Scheduling of CAES and Wind Power Generation in Joint Energy and reserve market: Toward More Realistic Results

Mohammad Ghaljehei[1,2,*], Mahrad Rahimi[1], Zahra Soltani[2], Behrouz Azimian[2], Behzad Vatandoust[3], Masoud Aliakbar Golkar[1]

[1] Faculty of Electrical Engineering, K. N. Toosi University of Technology, Tehran, Iran

[2] School of Electrical, Computer and Energy Engineering, Arizona State University, Tempe, AZ, US

[3] Electrical Power Engineering Unit, University of Mons, 7000 Mons, Belgium

* mghaljeh@asu.edu

**Abstract:** In this paper, a two-stage stochastic day-ahead (DA) scheduling model is proposed incorporating wind power units and compressed air energy storage (CAES) to clear a co-optimized energy and reserve market. The two-stage stochastic programming method is employed to deal with the wind power generation uncertain nature. A linearized AC optimal power flow (LAC-OPF) approach with consideration of network losses, reactive power, and voltage magnitude constraints is utilized in the proposed two-stage stochastic DA scheduling model. Using an engineering insight, a two-level LAC-OPF (TL-LAC-OPF) approach is proposed to (i) reduce the number of binary variables of the LAC-OPF approach which decreases the computational burden, and (ii) obtain LAC-OPF pre-defined parameters adaptively so that the accuracy of LAC-OPF approach is increased as a result of reducing artificial losses. Furthermore, as the CAES efficiency depends on its thermodynamic and operational conditions, the proposed two-stage stochastic DA scheduling model is developed by considering its thermodynamic characteristics to obtain a more realistic market decision at the first place. The proposed model is applied to IEEE 30-bus and 57-bus test systems using GAMS software, and is compared with three traditional approaches, i.e., AC-OPF, DC-OPF, and LAC-OPF. Simulation results demonstrate effectiveness of the proposed methodology.

## Nomenclature

Sets

| | |
|---|---|
| $N_G$ | Set of conventional thermal units |
| $N_F$ | Set of wind farms |
| $N_C$ | Set of compressed air energy storages (CAESs) |
| $N_B$ | Set of buses |
| $N_L$ | Set of lines |
| $N_\omega$ | Set of scenarios |
| $T$ | Total time intervals |

Indices

| | |
|---|---|
| $t$ | Index of time |
| $\omega$ | Index of scenario |
| $g$ | Index of conventional thermal unit |
| $f$ | Index of wind farm |
| $i, j$ | Index of bus |
| $k$ | Index of transmission line |
| $c$ | Index of CAES |

Parameters and variables

| | |
|---|---|
| $g_k$ | Real part of Matrix $Y_{Bus}$ for line $k$ |
| $b_k$ | Imaginary part of Matrix $Y_{Bus}$ for line $k$ |
| $b_{k0}$ | Shunt admittance of line $k$ |
| $\Delta V_i^{min}/\Delta V_i^{max}$ | Minimum/Maximum deviation of voltage magnitude of bus $i$ from 1 per-unit |
| $\theta^{max}$ | Maximum difference of voltage phase angle of two buses across a transmission line |
| $L$ | Number of linear blocks for linearizing loss |
| $AP_k^{max}$ | Maximum thermal MVA capacity of line $k$ |
| $R$ | Maximum number of linearized equations in thermal capacity constraint of transmission line |
| $P_{kt}^0/Q_{kt}^0$ | Active/Reactive power flow of line $k$ at time $t$ in first stage (day-ahead scheduling) |
| $PL_{kt}^0/QL_{kt}^0$ | Active/Reactive losses of line $k$ at time $t$ in first stage (day-ahead scheduling) |
| $\Delta V_{it}^0$ | Deviation of voltage magnitude of bus $i$ from 1 per-unit at time $t$ in first stage (day-ahead scheduling) |
| $\theta_{kt}^{0}$ | Difference of voltage phase angle of two buses across line $k$ at time $t$ in first stage (day-ahead scheduling) |
| $\Delta\theta_{kt}^{0}(l)$ | Linear block $l$ of angle difference in line $k$ at time $t$ in first stage (day-ahead scheduling) |
| $k^{0}(l)$ | Slope of linear block $l$ for linearizing loss in first stage (day-ahead scheduling) |
| $U_{ct}^{ChCAES}$ | Charging binary variable of CAES $c$ (1 if CAES $c$ is charged) |
| $U_{ct}^{DisCAES}$ | Discharging binary variable of CAES $c$ (1 if CAES $c$ is discharged) |
| $P_{c}^{ChCAES,max}/P_{c}^{ChCAES,min}$ | Maximum/minimum allowed charging power of CAES $c$ |
| $P_{c}^{DisCAES,max}/P_{c}^{DisCAES,min}$ | Maximum/minimum allowed discharging power of CAES $c$ |
| $P_{ct}^{ChCAES}/P_{ct}^{DisCAES}$ | Charging/Discharging power of CAES $c$ at time $t$ |
| $P_{ct}^{Ch}/P_{ct}^{Dis}$ | Total charging/discharging power capacity needed from CAES $c$ at time $t$ |
| $A_{c,t}$ | Stored air level of CAES $c$ at time $t$ |
| $AFR^{Ch}/AFR^{Dis}$ | Charging/Discharging airflow rate of CAES |
| $A_{ct}^{max}/A_{ct}^{min}$ | Maximum/minimum stored air level of CAES $c$ at time $t$ |
| $SUC_{g,t}$ | Start-up cost of thermal unit $g$ at time $t$ |
| $\underline{F_{g}}$ | Minimum cost of thermal unit $g$ for being committed with minimum power |
| $U_{g,t}$ | Binary variable (if thermal unit $g$ at time $t$ is on =1) |
| $P_{gtn}^{e}$ | Generating power of block $n$ in linearized fuel cost of thermal unit $g$ at time $t$ |
| $B_{gn}^{e}$ | Slope of block $n$ in linearized fuel cost of thermal unit $g$ |
| $C_{gt}^{DU}/C_{gt}^{DD}$ | Cost of up/down reserve capacity of thermal unit $g$ at time $t$ |
| $SR_{gt}^{usr}/SR_{gt}^{dsr}$ | Up/down reserve capacity of thermal unit $g$ at time $t$ |
| $C_{ct}^{Energy}$ | Cost of energy generating of CAES $c$ at time $t$ |
| $C_{ct}^{DU}/C_{ct}^{DD}$ | Cost of up/down reserve capacity of CAES $c$ at time $t$ |
| $SR_{ct}^{usrCAES}/SR_{ct}^{dsrCAES}$ | Up/down reserve capacity of CAES $c$ at time $t$ |
| $\Omega$ | Auxiliary variable, representative of expected cost of power balance in second stage |
| $\rho_{\omega}$ | Probability of scenario $\omega$ |
| $C_{gt}^{BU}/C_{gt}^{BD}$ | Cost of deployed up/down reserve capacity of thermal unit $g$ at time $t$ in second stage (real-time power balancing) |
| $\Delta r_{gt\omega}^{usr}/\Delta r_{gt\omega}^{dsr}$ | Deployed up/down reserve capacity of thermal unit $g$ at time $t$ in scenario $\omega$ |
| $C_{ct}^{BU}/C_{ct}^{BD}$ | Cost of deployed up/down reserve capacity of CAES $c$ at time $t$ |
| $\Delta r_{ct\omega}^{usrCAES}/\Delta r_{ct\omega}^{dsrCAES}$ | Deployed up/down reserve capacity of CAES $c$ at time $t$ in scenario $\omega$ in second stage (real-time power balancing) |
| $C_{ft}^{B,Spill}$ | Cost of wind power spillage of wind unit $f$ at time $t$ |
| $ws_{ft\omega}$ | Used power of wind unit $f$ at time $t$ in scenario $\omega$ in second stage (real-time power balancing) |
| $C_{it}^{B,VOLL}$ | Cost of mandatory load shedding at bus $i$ at time $t$ |
| $PILS_{it\omega}$ | Mandatory load shedding at bus $i$ at time $t$ in scenario $\omega$ in second stage (real-time power balancing) |
| $P_{gt}$ | Active power generation of thermal unit $g$ at time $t$ |
| $W_{ft}^{*}$ | Predicted power generation of wind unit $f$ at time $t$ |
| $PD_{it}/QD_{it}$ | Active/Reactive load of bus $i$ at time $t$ |
| $Q_{gt}^{0}$ | Reactive power generation of thermal unit $g$ at time $t$ in first stage (day-ahead scheduling) |
| $Y_{gt}$ | Start-up binary variable (if thermal unit $g$ at time $t$ is turned on =1) |
| $Z_{gt}$ | Shut-down binary variable (if thermal unit $g$ at time $t$ is turned off =1) |
| $MU_{g}/MD_{g}$ | Minimum up/down time of thermal unit $g$ |
| $P_{gn}^{e,max}$ | Maximum power of block $n$ in linearized fuel cost of thermal unit $g$ |
| $P_{g}^{max}/P_{g}^{min}$ | Maximum/minimum active power output of thermal unit $g$ |
| $Q_{g}^{max}/Q_{g}^{min}$ | Maximum/minimum reactive power output of thermal unit $g$ |
| $RU_{g}/RD_{g}$ | Ramp up/down rate of thermal unit $g$ |
| $\tau$ | Time resolution of the reserve capacity |
| $w_{ft\omega}$ | Wind power generation in scenario $\omega$ of wind unit $f$ at time $t$ in second stage (real-time power balancing) |
| $P_{kt\omega}/Q_{kt\omega}$ | Active/Reactive power flow of line $k$ at time $t$ in scenario $\omega$ in second stage (real-time power balancing) |
| $PL_{kt\omega}/QL_{kt\omega}$ | Active/Reactive losses of line $k$ at time $t$ in scenario $\omega$ in second stage (real-time power balancing) |
| $Q_{gt\omega}$ | Reactive power generation of thermal unit $g$ at time $t$ in scenario $\omega$ in second stage (real-time power balancing) |
| $\Delta V_{it\omega}$ | Deviation of voltage magnitude of bus $i$ from 1 per-unit at time $t$ in scenario $\omega$ in second stage (real-time power balancing) |
| $\theta_{kt\omega}$ | Difference of voltage phase angle of two buses across line $k$ at time $t$ in scenario $\omega$ in second stage (real-time power balancing) |
| $\Delta\theta_{kt\omega}(l)$ | Linear block $l$ of angle difference in line $k$ at time $t$ in scenario $\omega$ in second stage (real-time power balancing) |
| $k^{\omega}(l)$ | Slope of linear block $l$ for linearizing loss in scenario $\omega$ in second stage (real-time power balancing) |

## 1. Introduction

In recent years, wind power penetration has been significantly increasing due to its cost-effective and environment-friendly benefits [1]. However, its uncertain and variable power output leads to a need for greater operational flexibility to compensate for load-generation mismatch in real-time operation. The term "operational flexibility of power system" describes the ability to deal with variabilities and uncertainties while maintaining satisfactory reliability and security [2]. In the case of inadequate flexibility in scheduling decisions, a significant amount of wind power spillage (WPS) may occur in practice, which is not desirable with respect to the current sustainability policies. Traditionally, reserves provided by supply-side conventional resources are the majority of additional required reserves, which leads to higher losses in such units [3]. Apart from conventional power plants, energy storage systems (ESSs) can be considered as flexible options to contribute the operational flexibility of power systems [4]–[6]. However, including ESSs, such as compressed air energy storage (CAES), in the co-optimized energy and reserve market within the practical operational models such as stochastic AC network-constrained DA security-constrained unit commitment (SCUC) causes two significant challenges. First, detailed characteristics of the ESS should be included into the DA scheduling model to achieve a realistic SCUC decision for the co-optimized energy and reserve market. Second, the AC-based SCUC problem is a mixed-integer nonlinear programming (MINLP) problem, which incorporates non-convexity and nonlinearity, and no practical algorithm exists that is able to solve such complex problems to the global optimality [7], [8]. Hence, to address the mentioned challenges, the focus of this paper is to propose a tractable yet practical DA scheduling model incorporating wind power uncertainty and detailed operational model of CAES.

### 1.1. Literature Review

In [7], a stochastic SCUC along with network reconfiguration is presented to improve system flexibility. The model is developed based on a linearized AC-OPF (LAC-OPF) for the co-optimized clearing of energy and reserve market. In [9], a stochastic SCUC model has been suggested for the co-optimized energy and reserve market, where the network reconfiguration is utilized to improve system flexibility such that the additional cost incurred by wind power generation uncertainty is minimized. A multi-objective SCUC model based on AC optimal power flow (AC-OPF) is proposed in [10] for clearing a joint energy and reserve market. Minimizing joint market operational cost and maximizing steady-state stability of the system are the considered objective functions. However, since the model is based on AC-OPF which is a nonlinear non-convex problem, obtaining global optimality cannot be guaranteed. However, only conventional power plants are modeled in [7], [9], [10] without considering other flexible resources in the system.

Among all the energy storage units in large systems, the compressed air energy storage (CAES) and pumped hydro energy storage (PHES) have drawn more attention since they are able to discharge for a long period of time up to one day. Although PHES is a developed technology, further growth is limited by high initial investment cost, lack of suitable installation site, and environmental concerns [11]. The CAES system uses large natural reservoirs such as salt caverns, aquifers, and abandoned mines to store compressed air. The use of natural reservoirs will significantly decrease the investment cost of the CAES technology compared to others [12]. Moreover, long lifetime and low operation cost are other advantages of utilizing CAES system [13]. Authors in [14] have presented a stochastic self-scheduling model for a generation company considering CAES, thermal power plants, and renewable energy resources. In [12], a MINLP formulation is developed for a multi-objective problem, which minimizes the operation cost and maximizes the profit of scheduling wind power generation with CAES units in the energy market. The results of the study show that integrating the CAES units into the system reduces the operation cost by 6.7 percent and increases the profit by 43 percent; however, the uncertainty of wind power generation as well as transmission network constraints have not been considered. The uncertainty impact of wind power generation on day ahead scheduling with CAES units and thermal power plants was studied in [15], where a sensitivity analysis were conducted on the expected operation cost variation by changing the wind power penetration level and increasing the wind speed forecasting error. A stochastic multi-objective day-ahead scheduling model for a power system, which contains thermal generation units, plug-in electric vehicles, demand response programs, CAES units, and renewable distributed generations is proposed in [16]. However, the co-optimized energy and reserve market with CAES units was not evaluated in [12], [14]–[16]. In these references, a generic-model approach, which is used for all other energy storage systems, is utilized for CAES modeling; however, the lack of detailed CAES model, which describes the characteristics of this technology, needs to be incorporated in such studies. Also, the AC-based network constraints of transmission grid are not considered in [12], [14]–[16].

Reference [17] is one of the first studies on incorporating CAES units in the energy and reserve market, wherein the day-ahead SCUC problem is formulated for wind power generation and CAES units to participate in the energy and reserve market. A stochastic SCUC model was proposed in to study the integration of CAES and wind power units. Static voltage stability is considered in the proposed model to ensure the technical feasibility of the scheduling decision [4]. The proposed formulation in [4] and [17] that includes AC-based transmission network constraints make the model an MINLP problem. Besides, the uncertainty of wind power generation was not considered in [17]. Authors in [18] presented a model for clearing different US energy and reserve market with integration of CAES units, and the sensitivity of CAES net revenue to several design and performance parameters was investigated. However, the study is limited to CAES evaluation and does not propose a model for assessing the effect of CAES units in the grid from an independent system operator (ISO) point of view. References [3], [19]–[21] investigate the impact of the ideal and generic model of CAES on a joint energy and reserve market with uncertain wind power generation. For instance, a robust optimization scheduling framework was utilized in [3] to derive an optimal unit commitment decision in a system with high penetration of wind power incorporating demand response programs as well as bulk energy storage units in the co-optimized energy and reserve market. However, in all studies [3], [19]–[21], a DC-OPF is used to model the transmission network constraints. Furthermore, the ideal and generic model used for CAES does not represent the specific characteristics of this technology, which can result in unrealistic market decisions where CAES units may not be able to deliver their awarded reserves due to thermodynamic characteristics and consequently the system reliability can be jeopardized. Detailed thermodynamic characteristics of the CAES units in a self-scheduling model for the energy and reserve market were presented in [22], wherein a CAES unit, owned by a generation company, is scheduled in the energy and reserve market to obtain the maximum profit considering uncertain energy and reserve price. The study elaborated in [22] proves that considering the thermodynamic characteristics of CAES is of great importance, and applying the ideal and generic model leads to a considerable difference between real-world revenue and optimization output results.

### 1.2. Existing gaps and contributions

To the best of the authors knowledge, the following remarks on incorporating CAES in the energy and reserve market from ISO point of view can be extracted from the above literature survey, which require more attention and study:

- Most work [3], [9], [11]–[15], [18]–[21], [23]–[28] apply DC-OPF approach for transmission network constraints, since it leads to less computational complexity. However, DC-OPF assumptions create a gap between the output results of the AC and DC approaches. In some cases, the gap can be large enough so that the solution of stochastic SCUC based on DC-OPF may be unrealistic and jeopardize the reliability of an AC network [29]. Thus, the co-optimized energy and reserve market based on AC-OPF in the presence of CAES and wind power uncertainty in the stochastic SCUC problem has not been investigated sufficiently. A small number of studies [4], [10], and [17] has used a full AC-OPF in their energy and reserve market which make the market model hard to be solved.
- In the co-optimized energy and reserve market with CAES units, a generic CAES model, which is similar to any other energy storage system, is used in most of the work. The model is composed of constant charging/discharging efficiency, charging/discharging power limits at each time interval, and energy storage capacity limits. However, CAES efficiency is extremely dependent on its thermodynamic and operational conditions [22]. Thus, in order to obtain more realistic market outcomes, it is necessary to model the thermodynamic characteristics of CAES units in the stochastic DA SCUC problem and the co-optimized energy and reserve market.

Thus, in this paper, a novel two-stage stochastic DA scheduling model with linearized AC network constraints incorporating CAES and wind power uncertainty is proposed for clearing the co-optimized energy and reserve market. To achieve more realistic SCUC decisions and market outcomes, the CAES model includes its thermodynamic characteristics. Also, in order to cope with challenges of DC-OPF and full AC-OPF approaches, the LAC-OPF approach in [7] is utilized to solve the proposed stochastic two-stage DA scheduling model with AC transmission network constraints. However, the LAC-OPF approach in [7] imposes considerable number of binary variables on the proposed model which increases the computational burden of the problem. Also, as it will be proven, artificial losses (i.e., losses that are not real and are higher than the losses obtained from the AC power flow equations) may be created due to inaccurate selection of the pre-defined parameters (i.e., maximum difference of voltage phase angle of buses across a transmission line) needed in the LAC-OPF approach. In order to improve the proposed stochastic two-stage DA scheduling model based on LAC-OPF approach, an engineering insight is applied by developing a two-level linearized AC optimal power flow (TL-LAC-OPF) approach. In the first level, which is an offline process, the stochastic SCUC problem based on a simplified LAC-OPF approach can be solved to obtain (i) the sign of the flow in the transmission lines (that can determine the binary variables needed in the LAC-OPF problem), and (ii) the operating point of the transmission lines (that can determine the pre-defined parameters in the LAC-OPF problem). Then, the second level, which includes the stochastic SCUC problem with the full LAC-OPF approach (considering losses), is solved by fixing binary variables in the LAC-OPF based on the obtained sign of line flow (from the first offline level), and selecting accurate pre-defined parameters in the LAC-OPF based on the maximum difference of voltage phase angle of buses across a transmission line (from the first offline level). The aim of the proposed two-level model is to decrease the computational time of the optimization process and increase the accuracy of the LAC-OPF approach used in [7] for stochastic SCUC, and originally developed in [29].

The rest of the paper is organized as follows. Section 2 presents the problem formulation and proposed model. Section 3 discusses the simulation results, while Section 4 concludes the paper.

## 2. Problem formulation and Proposed methodology

### 2.1. Linearization of AC power flow

In order to increase the accuracy of the DC-OPF, a LAC-OPF approach is adopted in this paper with consideration of losses, voltage magnitude, and reactive power [29]. The LAC-OPF approach is based on the Taylor series and the assumptions that (i) the voltage magnitude of buses is roughly 1 per unit, (ii) the voltage angle difference between two buses in a transmission line is small, as $\sin(\theta_{kt}^{0}) \approx \theta_{kt}^{0}$ and $cos(\theta_{kt}^{0}) \approx 1$ [29]. The active and reactive power of line $k$ are written as (1) and (2).

$$P_{kt}^{0} = (V_{it}^{0})^2 g_k - V_{it}^{0} V_{jt}^{0} (g_k \cos\theta_{kt}^{0} + b_k sin\theta_{kt}^{0}) \quad (1)$$

$$Q_{kt}^{0} = -(V_{it}^{0})^2 (b_k + b_{k0}) + V_{it}^{0} V_{jt}^{0} (b_k \cos\theta_{kt}^{0} - g_k sin\theta_{kt}^{0}) \quad (2)$$

According to the first assumption, the voltage magnitude of bus $i$ can be written as (3), where $V_{it}^{0}$ is limited using (4).

$$V_{it}^{0} = 1 + \Delta V_{it}^{0} \quad (3)$$

$$\Delta V_i^{min} \le \Delta V_{it}^{0} \le \Delta V_i^{max} \quad (4)$$

By replacing (3) in (1) and (2), and neglecting second order terms and higher, (5) and (6) are obtained.

$$P_{kt}^{0} = (\Delta V_{it}^{0} - \Delta V_{jt}^{0}) g_k - b_k \theta_{kt}^{0} \quad (5)$$

$$Q_{kt}^{0} = -(1 + 2\Delta V_{it}^{0})(b_{k0}) - (\Delta V_{it}^{0} - \Delta V_{jt}^{0}) b_k - g_k \theta_{kt}^{0} \quad (6)$$

Furthermore, the active and reactive power loss of line $k$ based on the AC-OPF model can be approximated as (7) and (8), respectively [29].

$$PL_{kt}^{0} \approx g_k (\theta_{kt}^{0})^2 \quad (7)$$

$$QL_{kt}^{0} \approx b_k (\theta_{kt}^{0})^2 \quad (8)$$

Line losses in (7) and (8) are obtained using the second-order approximation of $cos\,\theta_{kt}^{0}$ and neglecting the higher orders. It should be noted that $(\theta_{kt}^{0})^2$ is a nonlinear term in these equations and can be linearized by applying piecewise linearization containing $L$ numbers of linear blocks as formulated in (9)-(16) [29].

$$(\theta_{kt}^{0})^2 \approx \sum_{l=1}^{L} k^0(l) \Delta\theta_{kt}^{0}(l) \quad (9)$$

where:

$$k^0(l) = (2l - 1)\theta^{max}/L \quad (10)$$

$$\sum_{l=1}^{L} \Delta\theta_{kt}^{0}(l) = |\theta_{kt}^{0}| = \theta_{kt}^{+,0} + \theta_{kt}^{-,0} \quad (11)$$

$$\theta_{kt}^{0} = \theta_{kt}^{+,0} - \theta_{kt}^{-,0} \quad (12)$$

$$0 \le \theta_{kt}^{+,0} \le \delta_{kt}^{0} \theta^{max} \quad (13)$$

$$0 \le \theta_{kt}^{-,0} \le (1 - \delta_{kt}^{0}) \theta^{max} \quad (14)$$

$$0 \le \Delta\theta_{kt}^{0}(l) \le \theta^{max}/L \quad (15)$$

$$\Delta\theta_{kt}^{0}(l) \le \Delta\theta_{kt}^{0}(l-1) \quad (16)$$

The slope of each linear block is obtained by (10). In (11) and (12), non-negative $\theta_{kt}^{-,0}$ and $\theta_{kt}^{+,0}$ variables are used as a replacement for $|\theta_{kt}^{0}|$ to calculate $\theta_{kt}^{0}$. Equations (13) and (14) enforce choosing one of $\theta_{kt}^{-,0}$ and $\theta_{kt}^{+,0}$ variables for each linear block and limiting the maximum value of these variables. The range limits of $\Delta\theta_{kt}^{0}(l)$ in each linear block is shown in (15). (16) guarantees the filling up of $\Delta\theta_{kt}^{0}(l)$ in the order of the number of linear blocks, as shown in Fig. 1.

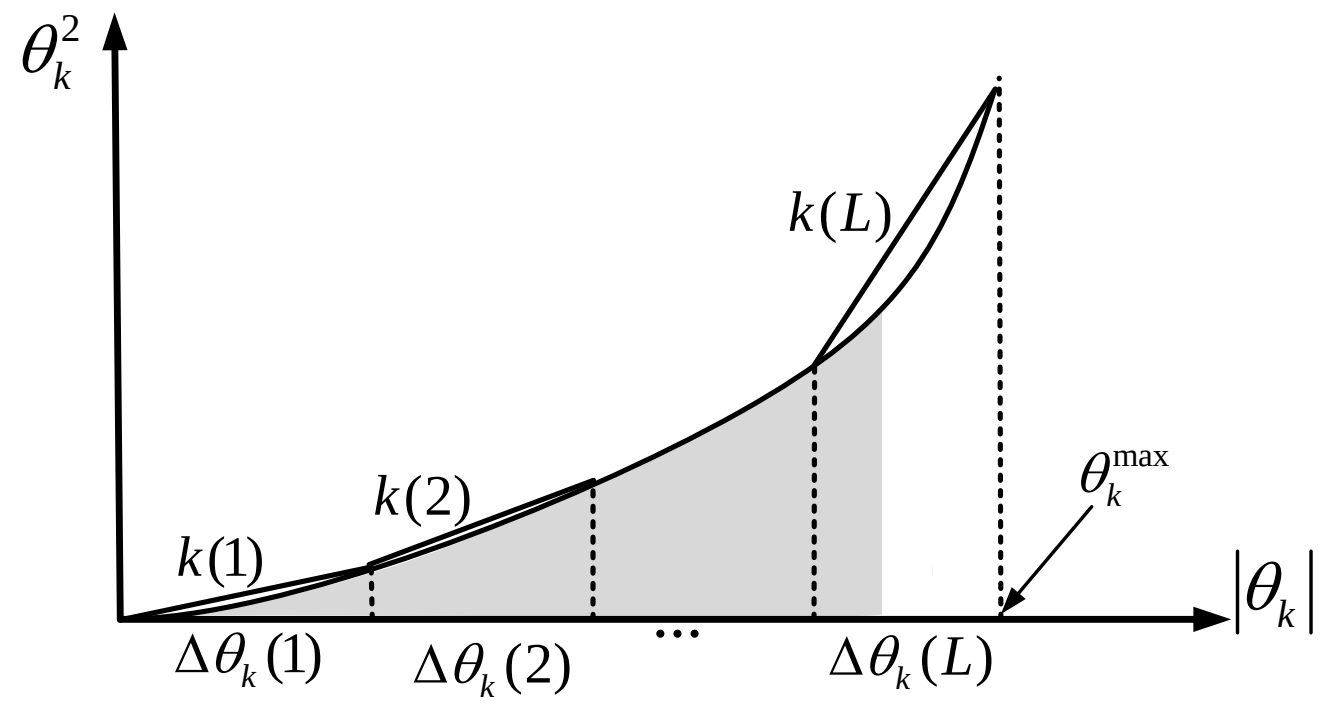

Fig. 1. Piecewise linearization of $\theta_k^2$.

Furthermore, the thermal capacity constraint of line $k$ is written as (17).

$$\left(P_{kt}^{0}\right)^2 + \left(Q_{kt}^{0}\right)^2 \leq (AP_k^{max})^2 \tag{17}$$

However, (17) is a nonlinear constraint, which can be linearized using (18) [30].

$$\left(sin\left(\frac{360°r}{R}\right) - sin\left(\frac{360°(r-1)}{R}\right)\right)P_{kt}^{0} - \left(sin\left(\frac{360°r}{R}\right) - sin\left(\frac{360°(r-1)}{R}\right)\right) \times Q_{kt}^{0} - AP_k^{max} sin\left(\frac{360°}{R}\right) \leq 0 \quad r = 1, \dots, R \tag{18}$$

Although the above LAC-OPF approach includes all essential variables of AC-OPF, it may cause two issues. First, the presented binary variables (i.e., $\delta_{kt}^{0}$) may increase the computation time of the MILP problem, especially in large-scale systems. For instance, for a network with $N_L$ as the number of transmission lines, in a 24-hour planning horizon, $N_L$×24 numbers of binary variables are added to the problem. It should be noted that the binary variable $\delta_{kt}^{0}$ is necessary to ensure that no additional artificial losses are imposed to the model [29]. Equations (12)-(14) illustrate that if the difference of voltage angles is positive, the value of $\delta_k$ is 1; otherwise, it is 0. This idea will be used to decrease the number of binary variables in next Section 2.4. Second, inaccurate selection of $\theta^{max}$ in the LAC-OPF approach may lead to inaccurate approximation of losses and consequently imposing artificial loss (i.e., losses that are not real and are higher than the losses obtained from the AC power flow equations) on the transmission lines. Fig. 2 illustrates the importance of determining an accurate value of $\theta^{max}$ in preventing artificial losses. As it can be seen in Fig. 2, for two linear blocks (i.e., $L = 2$), the value of linearized $\theta_k^2$ for an unnecessary high value of $\theta^{max}$ (blue in Fig. 2), is more than the corresponding value for a lower value of $\theta^{max}$ (red in Fig. 2) at an operating point. Therefore, based on (7) and (8), choosing same value for $\theta^{max}$ for all transmission lines which is not adaptive to the operating point of transmission lines can potentially lead to excessive artificial losses. In this paper, an offline procedure is proposed to address the aforementioned issues.

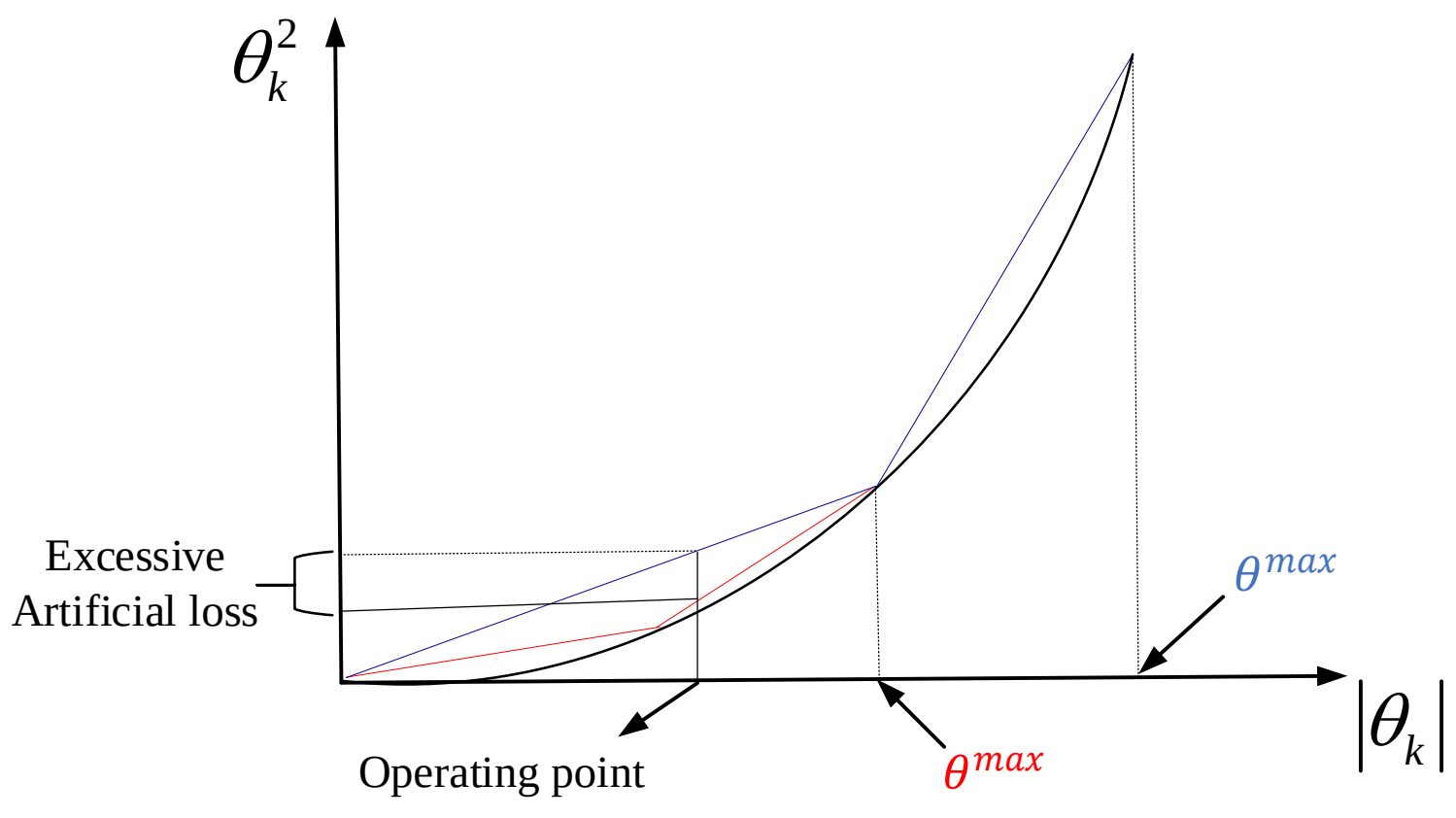

Fig. 2. Comparing linearized value of $\theta_k^2$ for different choice of $\theta^{max}$.

## 2.2. Thermodynamic model of CAES

The efficiency of CAES is defined by its energy ratio [31]. The energy ratio shows the amount of input electrical energy in per unit of output electrical energy [18], which is dependent on the output and input airflow rates. In the DA operational scheduling models with CAES [3], [11]–[15], [17]–[21], [26], [27], constant efficiency is considered. However, the airflow rates, and in turn, the efficiency of CAES are dependent on thermodynamic characteristics and operational conditions [22], [32]. For instance, the airflow rate depends on the amount of stored air in the reservoir during charging time, while during the discharging period, the airflow rate depends on the generated power. Therefore, to achieve more realistic results and preventing costly operation of the CAES units, the thermodynamic characteristics of CAES should be considered in the DA operational scheduling model for the co-optimized energy and reserve market. In the following, the model of CAES including its thermodynamic characteristics is presented for providing energy and reserve capacity in day-ahead power market.

### 2.2.1. Power capacity constraints

Constraint (19) prevents the CAES unit from charging and discharging at the same time. Constraints (20) and (21) are presented for limiting the charging and discharging powers of the CAES units. The total charging and discharging power capacity of CAES is expressed by (22) and (23).

$$U_{ct}^{ChCAES} + U_{ct}^{DisCAES} \leq 1 \quad (19)$$

$$U_{ct}^{ChCAES} P_{c}^{ChCAES,min} \leq P_{ct}^{ChCAES} + SR_{ct}^{dsrCAES} \leq U_{ct}^{ChCAES} P_{c}^{ChCAES,max} \quad (20)$$

$$U_{ct}^{DisCAES} P_{c}^{DisCAES,min} \leq P_{ct}^{DisCAES} + SR_{ct}^{usrCAES} \leq U_{ct}^{DisCAES} P_{c}^{DisCAES,max} \quad (21)$$

$$P_{ct}^{Ch} = P_{ct}^{ChCAES} + SR_{ct}^{dsrCAES} \quad (22)$$

$$P_{ct}^{Dis} = P_{ct}^{DisCAES} + SR_{ct}^{usrCAES} \quad (23)$$

### 2.2.2. Energy capacity constraints

The available energy in the CAES reservoir affects the airflow rate during charging [32]. In addition to this, the required airflow rate during discharge periods varies for different discharging power. Hence, considering these practical operational conditions, reservoir air capacity constraint can be written as (24) [22]:

$$A_{c,t+1} = A_{c,t} - \frac{P_{ct}^{Dis} \times AFR^{Dis}(P_{ct}^{Dis}) \times 3600}{CA^{max}} + \frac{P_{ct}^{Ch} \times AFR^{Ch}(A_{c,t}) \times 3600}{CA^{max}} \quad (24)$$

Based on (24), the amount of air that can be stored in the reservoir with power $P_{ct}^{Ch}$, is directly dependent on the available air level in the reservoir ($A_{c,t}$). Fig. 3 illustrates the effect of the available air level in the reservoir on the airflow rate during charging periods. According to this figure, as the stored air level increases, less air can be stored due to the air pressure in the reservoir. Also, according to (24), the required air level to be released from the reservoir to produce a specific amount of power is not constant, and directly depends on the level of generated power as shown in Fig.4. As it can be seen in Fig. 4, in lower generated power, higher airflow rate is required.

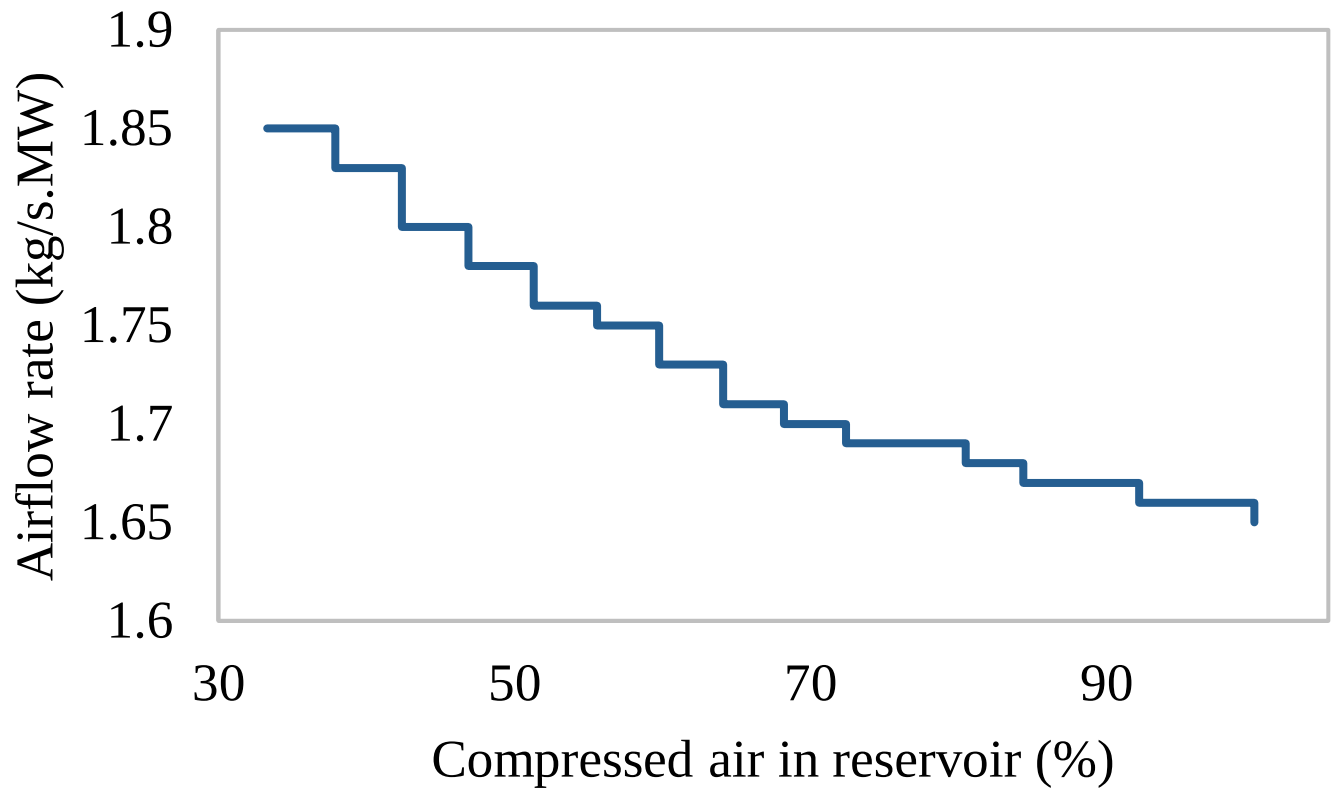

Fig. 3. Airflow rate versus the level of compressed air in the reservoir [22].

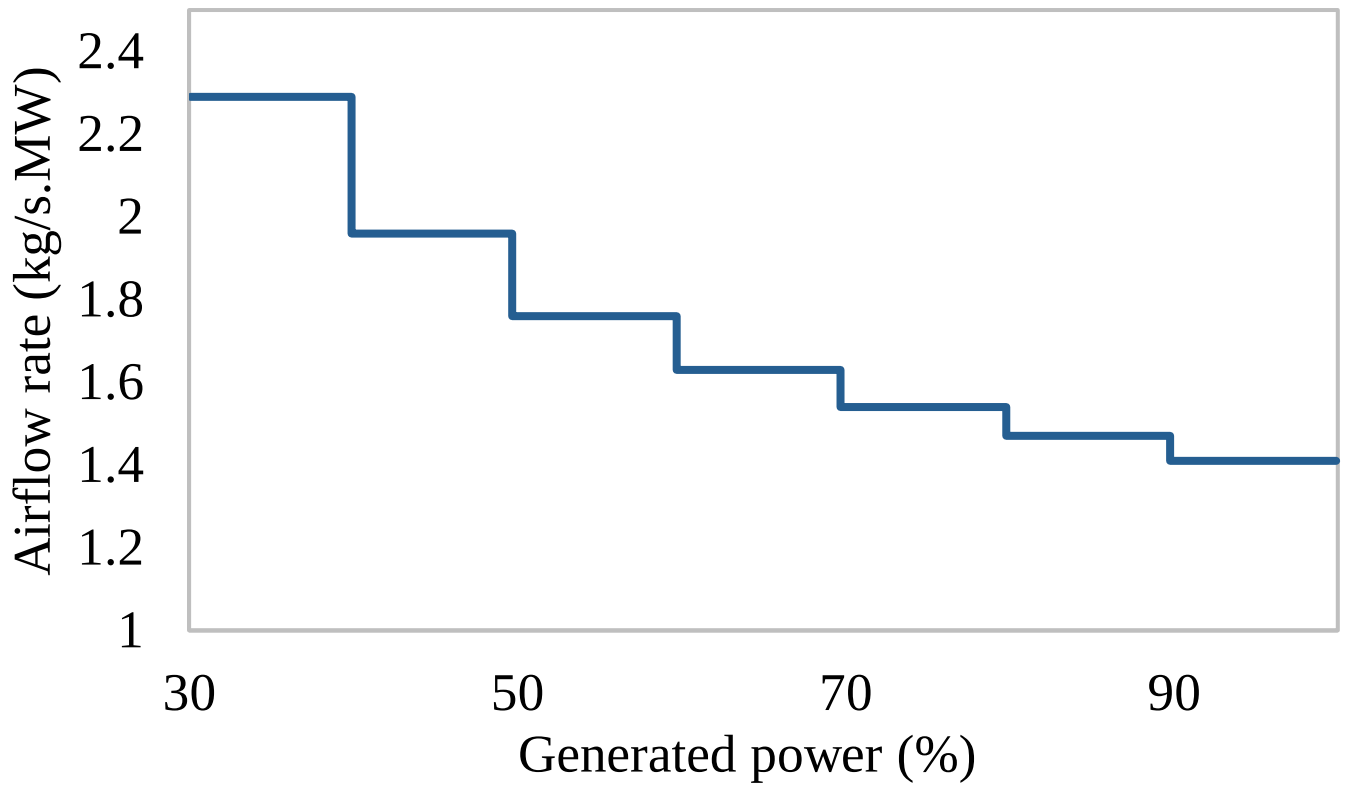

Fig. 4. Airflow rate versus generating power [22].

The term $P_{ct}^{Ch} \times AFR^{Ch}(A_{c,t})$ in (24) indicates the compressed and stored air, and the term $P_{ct}^{Dis} \times AFR^{Dis}(P_{ct}^{Dis})$ demonstrates the required amount of air released from the reservoir to produce $P_{ct}^{Dis}$. However, the integration of these terms into the optimization problem makes it nonlinear. In the following, the linearization of these terms is presented.

The curve in Fig. 3 can be modeled as a linearized curve with a couple of steps, as shown in Fig. 5 [22]. Based on the variables $b_{c,t,s}^{Ch}$ and $u_{c,t,s}^{Ch}$, and the parameters $A_{c,s}^{Ch,min}$ and $b_{c,s}^{Ch,max}$, equations (25)-(27) can be applied to choose the corresponding step of the compressed air in the reservoir.

$$A_{ct} = \sum_{s=1}^{M^{Ch}} \left[ b_{c,t,s}^{Ch} + u_{c,t,s}^{Ch} A_{c,s}^{Ch,min} \right] \quad (25)$$

$$0 \leq b_{c,t,s}^{Ch} \leq u_{c,t,s}^{Ch} b_{c,s}^{Ch,max} \quad (26)$$

$$\sum_{s=1}^{M^{Ch}} \left[ u_{c,t,s}^{Ch} \right] = 1 \quad (27)$$

where $A_{ct}$ is expressed linearly based on the variables $u_{c,t,s}^{Ch}$ and $b_{c,t,s}^{Ch}$. Equation (26) defines the limitation on each step of Fig. 5, and (27) guarantees that only one step is selected. The total compressed air when the compressor is operating at the power $P_{ct}^{Ch}$ is expressed as (28).

$$Air_{ct}^{Ch} = P_{ct}^{Ch} \times \sum_{s=1}^{M^{Ch}} u_{c,t,s}^{Ch} AFR_{s}^{Ch} \quad (28)$$

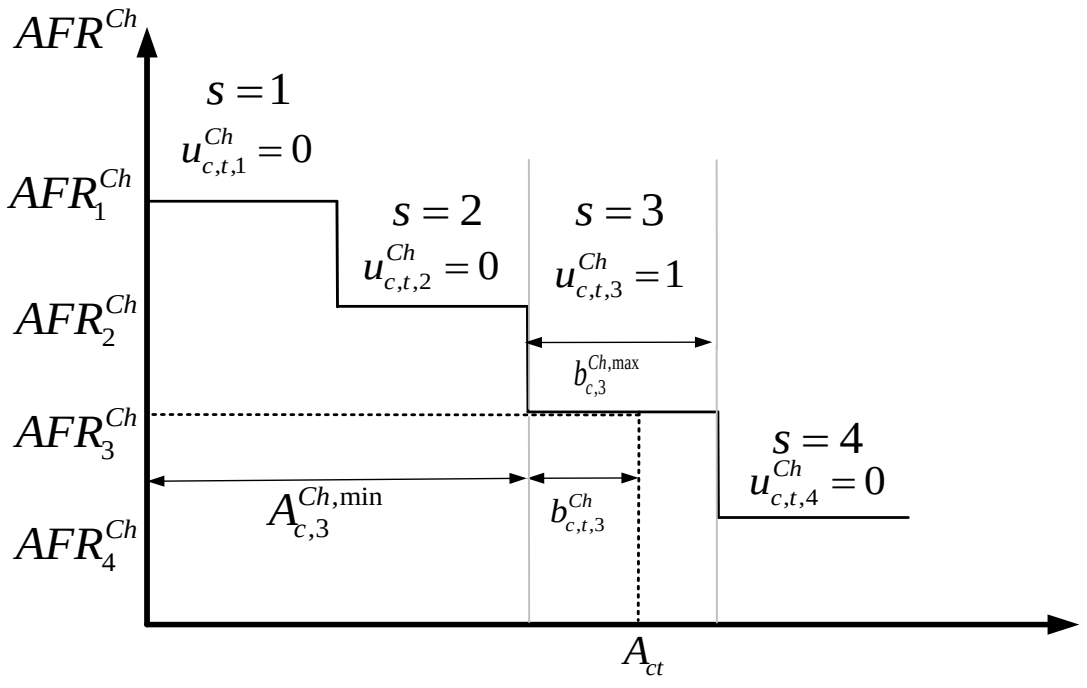

Fig. 5. Linearized curve of airflow rate versus the compressed air level in the reservoir.

However, (28) leads the problem to a nonlinear model due to the multiplication of $u_{c,t,s}^{Ch}$ and $P_{ct}^{Ch}$. Hence, (28) is linearized using big M method as can be seen in (29) and (30) [33].

$$AFR_{s}^{Ch} P_{ct}^{Ch} + M \times u_{c,t,s}^{Ch} \leq Air_{ct}^{Ch} + M \quad (29)$$

$$AFR_{s}^{Ch} P_{ct}^{Ch} - M \times u_{c,t,s}^{Ch} \geq Air_{ct}^{Ch} - M \quad (30)$$

Similarly, in order to linearize $P_{ct}^{Dis} \times AFR^{Dis}(P_{ct}^{Dis})$ in (24), the generated power can be formulated using linearized curve based on the discharging airflow rate as shown in Fig. 6 [22]. In Fig. 6, the variables $u_{c,t,s'}^{Dis}$ and $b_{c,t,s'}^{Dis}$ as well as the parameters $h_{c,s'}^{Dis,min}$ and $b_{c,s'}^{Dis,max}$ are considered for linearization of required airflow rate for generating various output powers. The linearized formulation can be expressed as (31)-(33).

$$P_{ct}^{Dis} = \sum_{s'=1}^{M^{Dis}} \left[ b_{c,t,s'}^{Dis} + u_{c,t,s'}^{Dis} h_{c,s'}^{Dis,min} \right] \tag{31}$$

$$0 \leq b_{c,t,s'}^{Dis} \leq u_{c,t,s'}^{Dis} b_{c,s'}^{Dis,max} \tag{32}$$

$$\sum_{s'=1}^{M^{Ch}} \left[ u_{c,t,s'}^{Dis} \right] = U_{ct}^{DisCAES} \tag{33}$$

In (31), the discharging power is expressed as a linear function of variables $u_{c,t,s'}^{Dis}$ and $b_{c,t,s'}^{Dis}$ while equation (32) defines the limitation on each step of Fig. 6. Equation (33) shows the relation between the discharging mode binary variable of CAES and the binary variables of each step. Finally, the total amount of required air for discharging the power $P_{ct}^{Dis}$ is expressed as (34).

$$Air_{ct}^{Dis} = \sum_{s'=1}^{M^{Dis}} \left( \left[ b_{c,t,s'}^{Dis} + u_{c,t,s'}^{Dis} h_{c,s'}^{Dis,min} \right] \times AFR_{s'}^{Dis} \right) \tag{34}$$

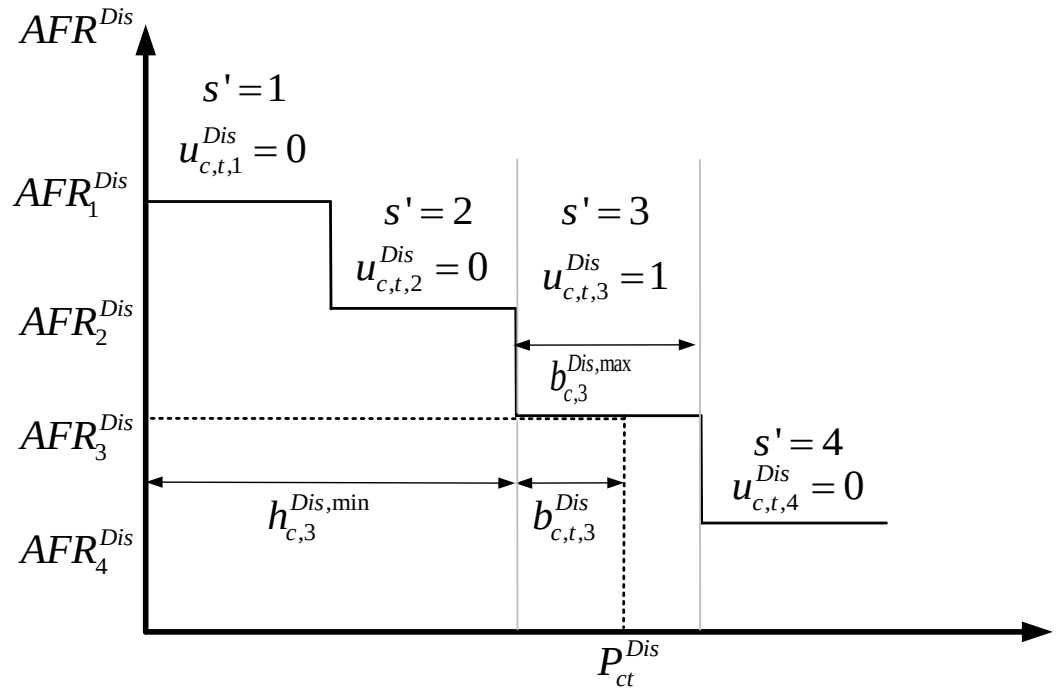

Fig. 6. Linearized curve of airflow rate versus generating power.

In order to complete the linearization process, the terms $Air_{ct}^{Dis}$ and $Air_{ct}^{Ch}$ obtained from the linearization process should replace the terms $P_{ct}^{Ch} \times AFR^{Ch}(A_{c,t})$ and $P_{ct}^{Dis} \times AFR^{Dis}(P_{ct}^{Dis})$ in (24), respectively, as shown in (35).

$$A_{c,t+1} = A_{c,t} - \frac{Air_{ct}^{Dis} \times 3600}{CA^{max}} + \frac{Air_{ct}^{Ch} \times 3600}{CA^{max}} \tag{35}$$

Other constraints of energy capacity including (36) and (37) limit the amount of compressed air in the reservoir and as the available compressed air at the beginning of scheduling (i.e., $t = 1$), respectively.

$$A_c^{min} \leq A_{c,t} \leq A_c^{max} \tag{36}$$

$$A_c^{initial} = B_c^{initial} A_c^{max} \tag{37}$$

### 2.3. Proposed two-stage stochastic DA scheduling model for clearing energy and reserve market considering wind resources and CAES

#### 2.3.1. Objective function

The objective function of the proposed model minimizes the cost of DA co-optimized energy and reserve market in a wind and CAES integrated power system from the ISO viewpoint. The proposed optimization problem is a two-stage optimization problem. In the first stage, the cost of the day-ahead electricity market including the energy and reserve market is minimized. In the second stage, the expected cost related to the power balance given different scenarios of wind power in real-time is minimized. It should be noted that the constraints of AC transmission network are considered at both stages.

$$\underset{\Xi_D}{Min} \sum_{t=1}^{T}\sum_{g=1}^{N_G}\left[SUC_{g,t} + \underline{F_g}.U_{g,t} + \sum_{n=1}^{NSF} P_{gtn}^{e} B_{gn}^{e}\right] + \sum_{t=1}^{T}\sum_{g=1}^{N_G}\left[C_{gt}^{DU} SR_{gt}^{usr} + C_{gt}^{DD} SR_{gt}^{dsr}\right] + \sum_{t=1}^{T}\sum_{c=1}^{N_C}\left[C_{ct}^{Energy} P_{ct}^{DisCAES} + C_{ct}^{DU} SR_{ct}^{usrCAES} + C_{ct}^{DD} SR_{ct}^{dsrCAES}\right] + \Omega \quad (38)$$

In (38), the first row includes the start-up cost of thermal power plants (the first term) and the piecewise linear function of quadratic fuel cost function [34] of these units (the second term). In the above objective function, the cost of wind power is neglected in the first stage and will be paid based on the marginal prices after clearing the market. The second and third row respectively indicate the up and down reserve capacity costs of thermal power plants, up and down reserve capacity costs of CAES units, and second stage cost denoted by $\Omega$ which is defined in (39).

$$\Omega = \sum_{\omega=1}^{N_\omega} \rho_\omega \begin{pmatrix} \sum_{t=1}^{T}\sum_{g=1}^{N_G}\left[C_{gt}^{BU} \Delta r_{gt\omega}^{usr} + C_{gt}^{BD} \Delta r_{gt\omega}^{dsr}\right] \\ \sum_{t=1}^{T}\sum_{c=1}^{N_C}\left[C_{ct}^{BU} \Delta r_{ct\omega}^{usrCAES} + C_{ct}^{BD} \Delta r_{ct\omega}^{dsrCAES}\right] \\ \sum_{t=1}^{T}\sum_{f=1}^{N_F}\left[C_{ft}^{B,Spill}(w_{ft\omega} - ws_{ft\omega})\right] + \sum_{t=1}^{T}\sum_{i=1}^{N_B}\left[C_{it}^{B,VOLL} PILS_{it\omega}\right] \end{pmatrix} \quad (39)$$

In the second stage, the objective function consists of deployment cost of up and down reserves of thermal power plants (the first row), the deployment cost of up and down reserves of the CAES units (the second row), and the WPS cost and involuntary load shedding cost (the third row).

### 2.3.2. Constraints of stochastic DA scheduling model for the first stage

Equations (41) and (42) are the active and reactive power balance constraints. The minimum up and down time of thermal power plants are expressed in (43) and (44).  Equations (45) and (46) are technical constraints for preventing simultaneous start-up and shut down of thermal units. Power generation constraints of thermal units are given as (47)-(52). Equations (53)-(54) are the reserve capacity constraints of thermal units.

(3)-(16), (18), (19)-(23), (25)-(27), and (29)-(37) (40)

$$\sum_{g\in i} P_{gt} + \sum_{c\in i}(P_{ct}^{DisCAES} - P_{ct}^{ChCAES}) + \sum_{f\in i} W_{ft}^{*} + \sum_{k\in i} P_{kt}^{0} + \sum_{k\in i} 0.5PL_{kt}^{0} = PD_{it} \quad (41)$$

$$\sum_{g\in i} Q_{gt}^{0} + \sum_{k\in i} Q_{kt}^{0} - \sum_{k\in i} 0.5QL_{kt}^{0} = QD_{it} \quad (42)$$

$$Y_{gt} + \sum_{n=1}^{MU_g-1} Z_{g,t+n} \le 1 \quad (43)$$

$$Z_{gt} + \sum_{n=1}^{MD_g-1} Y_{g,t+n} \le 1 \quad (44)$$

$$Y_{gt} - Z_{gt} = U_{gt} - U_{g,t-1} \quad (45)$$

$$Y_{gt} + Z_{gt} \le 1 \quad (46)$$

$$P_{gt} = P_g^{min}.U_{g,t} + \sum_{n=1}^{NSF} P_{gtn}^{e} \quad (47)$$

$$0 \le P_{gtn}^{e} \le P_{gn}^{e,max} \quad (48)$$

$$U_{gt} P_g^{min} \le P_{g,t} \le U_{gt} P_g^{max} \quad (49)$$

$$U_{gt}Q_g^{min} \le Q_{g,t}^0 \le U_{gt}Q_g^{max} \tag{50}$$

$$P_{gt} + SR_{gt}^{usr} \le U_{gt}P_g^{max} \tag{51}$$

$$P_{gt} - SR_{gt}^{dsr} \ge U_{gt}P_g^{min} \tag{52}$$

$$0 \le SR_{gt}^{usr} \le U_{gt}RU_g\tau \tag{53}$$

$$0 \le SR_{gt}^{dsr} \le U_{gt}RD_g\tau \tag{54}$$

**2.3.3. Constraints of DA scheduling model for the second stage**

The constraints of transmission network based on LAC-OPF in the second stage are given as (55)-(69). Constraint (70) is for the involuntary load shedding limit. The constraint of WPS can be expressed as (71). The constraints of the deployed spinning reserve of thermal power plants and CAES units are given as (72)-(75). The constraints of the ramp rate limits of thermal units are described as (76) and (77) [35].

$$\sum_{g\in i}\left(\Delta r_{gt\omega}^{usr} - \Delta r_{gt\omega}^{dsr}\right) + \sum_{c\in i}\left(\Delta r_{ct\omega}^{usrCAES} - \Delta r_{ct\omega}^{dsrCAES}\right) + \sum_{f\in i}\left(w_{ft\omega} - W_{ft}^{*} - ws_{ft\omega}\right) + \sum_{k\in i}\left(P_{kt\omega} - P_{kt}^0\right) + \sum_{k\in i} 0.5\left(PL_{kt\omega} - PL_{kt}^0\right) = -PILS_{it\omega} \tag{55}$$

$$\sum_{g\in i} Q_{gt\omega} + \sum_{k\in i}\left(Q_{kt\omega} - Q_{kt}^0\right) - \sum_{k\in i} 0.5\left(QL_{kt\omega} - QL_{kt}^0\right) = -QILS_{it\omega} \tag{56}$$

$$QILS_{it\omega} \times PD_{it} = QD_{it} \times PILS_{it\omega} \tag{57}$$

$$P_{kt\omega} = (\Delta V_{it\omega} - \Delta V_{jt\omega})g_k - b_k\theta_{kt\omega} \tag{58}$$

$$Q_{kt\omega} = -(1 + 2\Delta V_{it\omega})(b_{k0}) - (\Delta V_{it\omega} - \Delta V_{jt\omega})b_k - g_k\theta_{kt\omega} \tag{59}$$

$$PL_{kt\omega} = g_k \sum_{l=1}^{L} k^{\omega}(l)\Delta\theta_{kt\omega}(l) \tag{60}$$

$$QL_{kt\omega} = b_k \sum_{l=1}^{L} k^{\omega}(l)\Delta\theta_{kt\omega}(l) \tag{61}$$

$$k^{\omega}(l) = (2l - 1)\theta^{max}/L \tag{62}$$

$$\sum_{l=1}^{L} \Delta\theta_{kt\omega}(l) = |\theta_{kt\omega}| = \theta_{kt\omega}^{+} + \theta_{kt\omega}^{-} \tag{63}$$

$$\theta_{kt\omega} = \theta_{kt\omega}^{+} - \theta_{kt\omega}^{-} \tag{64}$$

$$0 \le \theta_{kt\omega}^{+} \le \delta_{kt\omega}\theta^{max} \tag{65}$$

$$0 \le \theta_{kt\omega}^{-} \le (1 - \delta_{kt\omega})\theta^{max} \tag{66}$$

$$0 \le \Delta\theta_{kt\omega}(l) \le \theta^{max}/L \tag{67}$$

$$\Delta\theta_{kt\omega}(l) \le \Delta\theta_{kt\omega}(l - 1) \tag{68}$$

$$\left(sin\left(\frac{360°r}{R}\right) - sin\left(\frac{360°(r-1)}{R}\right)\right)P_{kt\omega} - \left(sin\left(\frac{360°r}{R}\right) - sin\left(\frac{360°(r-1)}{R}\right)\right) \times Q_{kt\omega} - AP_k^{max} sin\left(\frac{360°}{R}\right) \le 0 \quad r = 1, \dots, R \tag{69}$$

$$0 \le PILS_{it\omega} \le PD_{it} \tag{70}$$

$$0 \le ws_{ft\omega} \le w_{ft\omega} \tag{71}$$

$$0 \le \Delta r_{gt\omega}^{usr} \le SR_{gt}^{usr} \tag{72}$$

$$0 \le \Delta r_{gt\omega}^{dsr} \le SR_{gt}^{dsr} \tag{73}$$

$$0 \le \Delta r_{ct\omega}^{dsrCAES} \le SR_{ct}^{dsrCAES} \tag{74}$$

$$0 \le \Delta r_{ct\omega}^{usrCAES} \le SR_{ct}^{usrCAES} \tag{75}$$

$$P_{g,t+1} + \Delta r_{g,t+1,\omega}^{usr} - \Delta r_{g,t+1,\omega}^{dsr} - P_{g,t} - \Delta r_{gt\omega}^{usr} + \Delta r_{gt\omega}^{dsr} \le RU_g \tag{76}$$

$$P_{g,t-1} + \Delta r^{usr}_{g,t-1,\omega} - \Delta r^{dsr}_{g,t-1,\omega} - P_{g,t} - \Delta r^{usr}_{gt\omega} + \Delta r^{dsr}_{gt\omega} \leq RD_g \quad (77)$$

As discussed, although this LAC-OPF-based model includes all essential variables of the AC-OPF model, the presented binary variables for loss linearization ($\delta_{kt\omega}$ and $\delta^0_{kt}$) may increase the computation time of MILP model. In this model, $N_L \times (1 + N_\omega) \times 24$ binary variables are added to the MILP problem for a network with $N_L$ transmission lines in a 24-hour scheduling horizon. Also, non-adaptive selection of the parameters $\theta^{max}$ (i.e., same $\theta^{max}$ for all the transmission lines in the system) can potentially create artificial losses for the transmission lines as discussed in Section 2.1. In the following, based on an engineering insight, a two-level LAC-OPF approach, where the first stage is an offline process, while the second stage it the original DA scheduling model, is presented to tackle the above two problems.

### 2.4. Proposed two-level two-stage stochastic DA scheduling model for clearing energy and reserve market considering wind resources and CAES

In the proposed two-stage stochastic DA scheduling model presented in Section 2.3, the binary variables in the LAC-OPF approach indicate the sign of voltage phase angle difference in the transmission lines. If these signs can be obtained before solving the problem, the number of binary variables of the proposed MILP model can be reduced, which in turn decreases the solving time and complexity of the MILP problem. Since the transmission line losses are insignificant compared to the transmitted power for supplying system loads, the losses rarely affect the sign of voltage phase angle difference of the buses. Hence, using this engineering insight, a two-level LAC-OPF approach is proposed in this paper. In the first level, which is an offline process, the stochastic SCUC problem based on the LAC-OPF approach can be solved without considering transmission line losses to obtain (i) the sign of the flow in the transmission lines, and (ii) the operating point of the transmission lines to determine the difference of voltage phase angle of two buses across a transmission line for each time and scenario. Then, the second level, which includes the stochastic SCUC problem with the full LAC-OPF approach (considering losses), is solved by using the sign of line flow to fix binary variables in the LAC-OPF and the maximum difference of voltage phase angle of buses across the transmission lines, obtained from the first offline level. In other words, in the second level, the binary variables required for calculating the sign of voltage phase angle difference are changed into the parameters obtained from the first offline level. Consequently, the number of binary variables of the proposed MILP model reduces. In next Section, the proposed two-level two-stage stochastic DA scheduling model and corresponding constraints are presented.

## 2.5. First level of the proposed two-level two-stage stochastic DA scheduling model for clearing energy and reserve market considering wind resources and CAES

The purpose of this level is to obtain the binary variables of LAC-OPF and the maximum difference of voltage phase angle of buses across the transmission lines by solving the two-stage stochastic DA scheduling model presented in Section 2.3 without considering the losses.

### 2.5.1.1. Objective function

The objective function of the first level is the same as the objective function of the single-level two-stage stochastic DA scheduling problem presented in Section 2.3, as can be seen in (38). The objective function for the first level (i.e., offline) should be minimized subject to the following constraints.

### 2.5.1.2. Constraints of the proposed stochastic DA scheduling model at the first level

The constraints of the proposed model in the first level include (78)-(82), where the constraints of power balance constraint based on LAC-OPF in the first and second stages are given in (79)-(80) and (81)-(82), respectively.

(40), (43)-(54), and (57)-(77) (78)

$$\sum_{g\in i} P_{gt} + \sum_{c\in i}\left(P_{ct}^{DisCAES} - P_{ct}^{ChCAES}\right) + \sum_{f\in i} W_{ft}^{*} + \sum_{k\in i} P_{kt}^{0} = PD_{it} \tag{79}$$

$$\sum_{g\in i} Q_{gt}^{0} + \sum_{k\in i} Q_{kt}^{0} = QD_{it} \tag{80}$$

$$\sum_{g\in i}\left(\Delta r_{gt\omega}^{usr} - \Delta r_{gt\omega}^{dsr}\right) + \sum_{g\in i}\left(\Delta r_{ct\omega}^{usrCAES} - \Delta r_{ct\omega}^{dsrCAES}\right) + \sum_{f\in i}\left(w_{ft\omega} - W_{ft}^{*} - ws_{ft\omega}\right) + \sum_{k\in i}\left(P_{kt\omega} - P_{kt}^{0}\right) = -PILS_{it\omega} \tag{81}$$

$$\sum_{g\in i} Q_{gt\omega} + \sum_{k\in i}\left(Q_{kt\omega} - Q_{kt}^{0}\right) = -QILS_{it\omega} \tag{82}$$

## 2.6. Second level of the proposed two-level two-stage stochastic DA scheduling model for clearing energy and reserve market considering wind resources and CAES

In second level, the LAC-OPF-based stochastic SCUC problem is solved including losses using the line flow signs to fixed $\delta_{kt\omega}$ and $\delta_{kt}^{0}$ and to determine adaptive maximum difference of voltage phase angle of buses across the transmission lines (i.e., $\theta_{kt}^{max,0}$ and $\theta_{kt\omega}^{max}$).

### 2.6.1.1. Objective function

The objective function of the second level is the same as the objective function of the single-level two-stage stochastic DA scheduling problem presented in Section 2.3, as can be seen in (38). The objective function for the second level should be minimized subject to the following constraints.

2.6.1.2. Constraints of the proposed stochastic DA scheduling model at the second level

The constraints of the proposed model in the second level include (83)-(92), where $\delta_{kt\omega}$ and $\delta_{kt}^{0}$ are obtained from the first offline level, and the $\theta_{kt}^{max,0}$ and $\theta_{kt\omega}^{max}$ are adaptive to each specific transmission line operating point.

(3)-(9), (11)-(12), (16), (18), (19)-(23), (25)-(27), (29)-(37), (41)-(61), (63)-(64), and (68)-(77) (83)

$$k^{0}(l) = (2l-1)\theta_{kt}^{max,0}/L \quad (84)$$

$$0 \le \theta_{kt}^{+,0} \le \delta_{kt}^{0}\theta_{kt}^{max,0} \quad (85)$$

$$0 \le \theta_{kt}^{-,0} \le (1-\delta_{kt}^{0})\theta_{kt}^{max,0} \quad (86)$$

$$0 \le \Delta\theta_{kt}^{0}(l) \le \theta_{kt}^{max,0}/L \quad (87)$$

$$k^{\omega}(l) = (2l-1)\theta_{kt\omega}^{max}/L \quad (88)$$

$$0 \le \theta_{kt\omega}^{+} \le \delta_{kt\omega}\theta_{kt\omega}^{max} \quad (89)$$

$$0 \le \theta_{kt\omega}^{-} \le (1-\delta_{kt\omega})\theta_{kt\omega}^{max} \quad (90)$$

$$0 \le \Delta\theta_{kt\omega}(l) \le \theta_{kt\omega}^{max}/L \quad (91)$$

Fig. 7 illustrates the framework of the proposed two-level two-stage stochastic DA scheduling model for the co-optimized energy and reserve market in the presence of wind power and CAES.

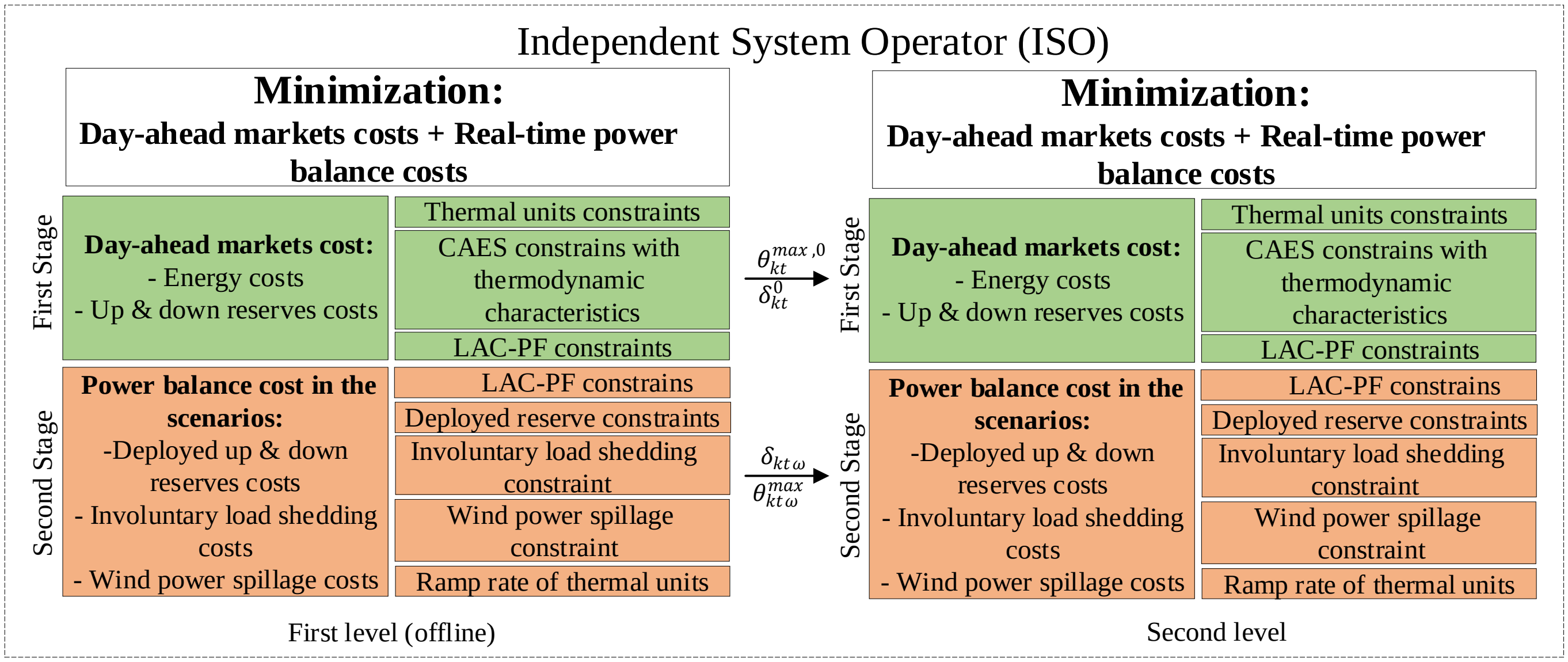

Fig. 7. Framework of proposed two-level two-stage stochastic DA scheduling in the presence of wind power and CAES.

## 3. Simulation Results

The proposed two-level two-stage stochastic DA scheduling model for the co-optimized energy and reserve market in the presence of wind power and CAES has been implemented by a computer with Intel Core i5 2.0 GHz processor and 4 GB RAM using GAMS software [36]. Simulations are divided into two studies A and B. In study A, the proposed two-level approach is used for solving the LAC-OPF-based SCUC problem on IEEE 30-bus and IEEE 57-bus systems, without considering wind farms and CAES.

Then, the results of the proposed two-level approach are compared with three other methods to prove its effectiveness. In study B, the results of the proposed two-stage two-level stochastic DA scheduling model for the co-optimized energy and reserve market in the presence of wind farms and CAES are presented. Also, the effects of CAES and its thermodynamic model on the DA outcomes are investigated.

#### 3.1.1. Study A

For solving the deterministic SCUC problem, the two-level linearized AC optimal power flow (TL-LACOPF) is applied to IEEE 30-bus and 57-bus systems for a 24-hour scheduling herizon. Then, the results are compared with the results of other available optimal power flow models, such as DC power flow (DC-PF) [3], full AC power flow (AC-OPF) [17], and linearized AC power flow (LACPF) approaches [29]. It should be noted that based on TL-LAC-OPF, LAC-OPF, and DC-PF approaches, the SCUC problem can be formulated as a MILP model, while based on AC-OPF approach, this problem is formulated as an MINLP model. Please note that since the SCUC model based on the AC-OPF is an MINLP problem, the existing solvers only can give a feasible solution (non-optimal) for it. The number of linear blocks for the cost function of thermal units is considered 5. Moreover, the CPLEX solver is used to solve the MILP models [37]. In the following, the results of these four methods for the two mentioned systems are demonstrated.

##### 3.1.1.1. IEEE-30 bus system

Table I shows the characteristic of thermal units. Besides, Fig. 8 illustrates the hourly load at each bus. The total operation costs of four models are compared in Table II. As expected, due to the weakness of the presented algorithms to solve complex MINLP-based problems, the results of the AC-OPF-based SCUC problem have the highest cost among other methods. This cost is considered as a reference to compare the cost reduction efficiency of other models. In the DC-PF model, the SCUC problem has the lowest cost, $133469.7, which is $7515 less than the cost of AC-OPF model. The reason is that the limitations of AC transmission network, including voltage constraints, reactive power, and active and reactive losses of transmission lines, have been ignored. Furthermore, because of the model simplicity, the time required to reach the optimal answer is very short, i.e., 1.2 seconds. The total cost of operation for the LAC-OPF approach is $137651.4, with simulation time of 1240.2 seconds. The cost difference between this model and the MINLP model is $3333.2. Regarding the TL-LAC-OPF approach, the operation cost, the cost reduction compared to the AC-OPF model, and the simulation time are $137013.0, $3971.7, and 35.6 seconds, respectively. As it can be seen, the TL-LAC-OPF approach has a lower operation costs with much shorter time, compared to the LAC-OPF approach. The reason for this cost reduction is the reduced error in the linearized model due to the more adaptive selection of the maximum difference of voltage phase angle of buses across a transmission line which prevents artificial losses, and this results in reduction in the operation costs (this error will be discussed in Table V). On the other hand, owing to be a two-level model, the complexity of the MILP model is decreased, which reduces the simulation time of TL-LAC-OPF approach compared to LAC-OPF approach (roughly 35 times faster).

Table I. Thermal unit characteristics in IEEE 30-bus system.

| Unit # | SU | RD | RU | MD | MU | $c_g$ | $b_g$ | $a_g$ | $Q^{min}$ | $Q^{max}$ | $P^{min}$ | $P^{max}$ |
|---|---|---|---|---|---|---|---|---|---|---|---|---|
| 1 | 440 | 50 | 50 | 10 | 10 | 6.8 | 20 | 0.038 | -20 | 150 | 10 | 100 |
| 2 | 440 | 50 | 50 | 10 | 10 | 6.8 | 20 | 0.25 | -20 | 60 | 10 | 80 |
| 3 | 100 | 25 | 25 | 1 | 1 | 36 | 42 | 0.15 | -15 | 40 | 10 | 60 |
| 4 | 100 | 30 | 30 | 1 | 1 | 36 | 42 | 0.15 | -15 | 47.8 | 10 | 65 |
| 5 | 100 | 25 | 25 | 1 | 1 | 36 | 42 | 0.15 | -15 | 44.7 | 10 | 70 |
| 6 | 100 | 30 | 30 | 1 | 1 | 36 | 42 | 0.15 | -15 | 62.5 | 10 | 60 |

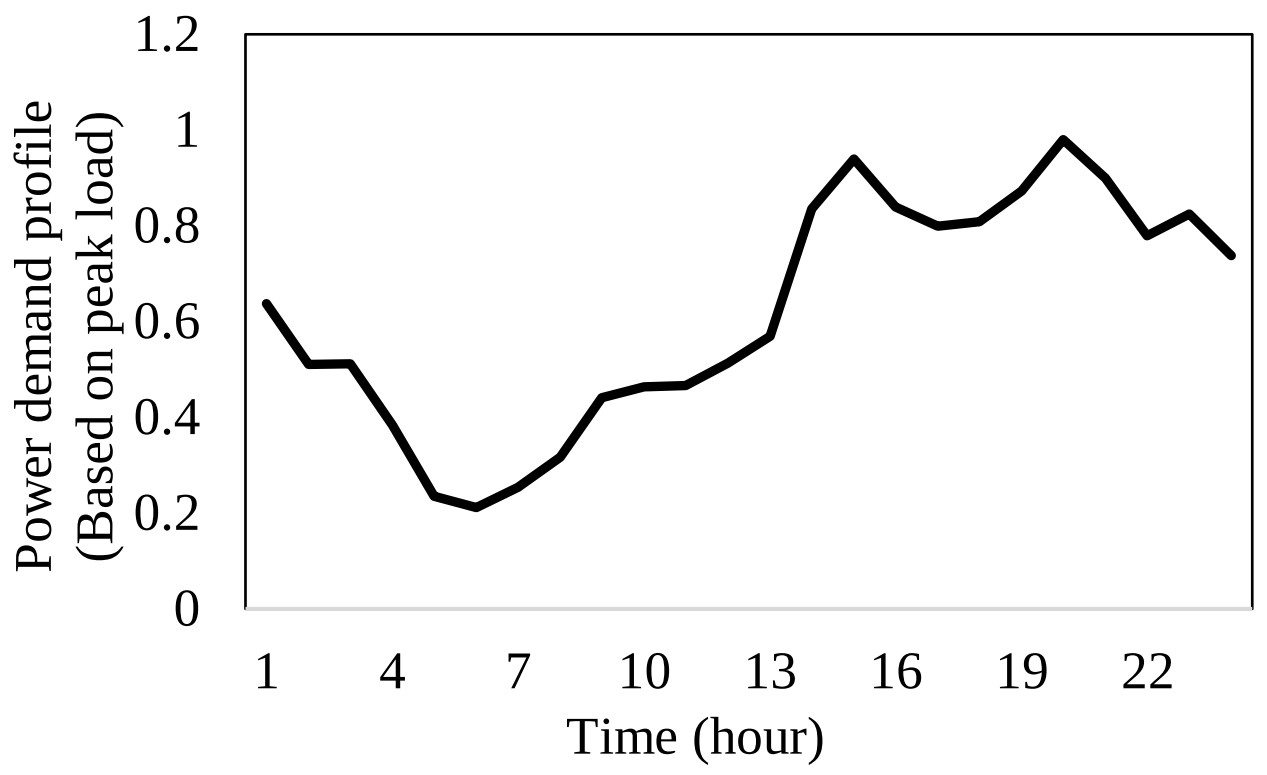

Fig. 8. Hourly load at each bus for the IEEE 30-bus system.

Table II. The comparison of simulation results obtained for four models in the IEEE 30-bus system.

| | AC-OPF Model | TL-LACPF Model | LACPF Model | DC-PF Model |
|---|---|---|---|---|
| Total operating cost ($) | 140984.7 | 137013.0 | 137651.4 | 133469.7 |
| Simulation Time (sec.) | 253.8 | 35.6 | 1240.2 | 1.2 |

### 3.1.1.2. IEEE 57-bus system

To investigate the TL-LAC-OPF approach performance for larger systems, it is also applied to the IEEE 57-bus system, and the results are compared with three other models. The single line diagram of this system with transmission lines parameters are accessible in [38] and [4]. Fig. 9 shows the hourly load at each bus, based on the corresponding maximum load. Table III presents the characteristics of thermal units [4]. Table IV demonstrates the simulation results of the IEEE 57-bus system. It is evident from table IV that the operation cost of the AC-OPF model is higher than other models, similar to the IEEE 30-bus system. Moreover, the DC-PF model has the lowest cost. The operation cost of the TL-LAC-OPF approach is $599438.1, which is $2642.9 lower than the cost of the LAC-OPF approach, i.e., $602081.0. Comparing the cost reduction of the LAC-OPF and TL-LAC-OPF approaches with the MINLP model

shows that the operation cost of the TL-LAC-OPF approach has experienced a greater decrease. More clearly, in the TL-LAC-OPF approach, the adaptive selection of the maximum difference of voltage phase angle of buses across a transmission line reduces excessive artificial losses, and it results in costs reduction.

Table III. Specifications of thermal units for IEEE 57-bus system

| Unit # | SU | RD | RU | MD | MU | $c_g$ | $b_g$ | $a_g$ | $Q^{min}$ | $Q^{max}$ | $P^{min}$ | $P^{max}$ |
|---|---|---|---|---|---|---|---|---|---|---|---|---|
| 1 | 176 | 120 | 120 | 2 | 3 | 0 | 20 | 0.0755 | -140 | 200 | 50 | 575.88 |
| 2 | 187 | 50 | 50 | 1 | 3 | 0 | 40 | 0.01 | -17 | 50 | 10 | 100 |
| 3 | 113 | 50 | 50 | 1 | 2 | 0 | 20 | 0.25 | -10 | 60 | 20 | 140 |
| 4 | 267 | 50 | 50 | 2 | 4 | 0 | 40 | 0.01 | -8 | 25 | 10 | 100 |
| 5 | 180 | 350 | 350 | 1 | 1 | 0 | 20 | 0.0222 | -140 | 200 | 40 | 550 |
| 6 | 113 | 25 | 25 | 1 | 1 | 0 | 40 | 0.01 | -3 | 9 | 10 | 100 |
| 7 | 176 | 105 | 105 | 1 | 2 | 0 | 20 | 0.0322 | -150 | 155 | 30 | 410 |

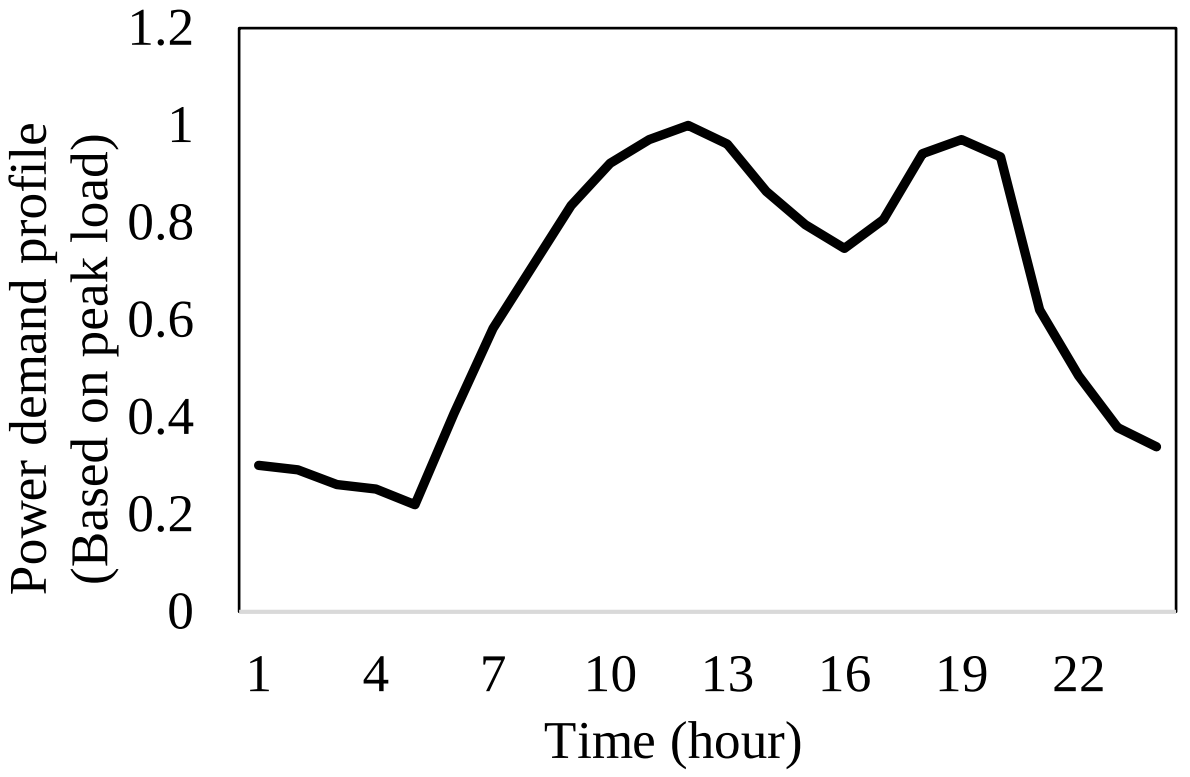

Fig. 9. Hourly load at each bus for IEEE 57-bus system.

Table IV. The comparison of simulation results obtained for four models in IEEE 57-bus system

| | AC-OPF approach | TL-LAC-OPF approach | LAC-OPF approach | DC-PF approach |
|---|---|---|---|---|
| Total operating cost ($) | 626115.4 | 599438.1 | 602081.0 | 590295.8 |
| Simulation Time (sec.) | 429.2 | 61.8 | 2346.7 | 3.8 |

Table V. Error comparison of LAC-OPF and TL-LAC-OPF resulting in excessive artificial losses (%)

| | IEEE 30-bus system | IEEE 57-bus system |
|---|---|---|
| TL-LAC-OPF approach | 1.8 | 0.8 |
| LAC-OPF approach | 33.9 | 37.7 |

Table V presents the error of active losses between the linear and nonlinear models for LAC-OPF and TL-LAC-OPF. It should be noted that the nonlinear losses are calculated based on the results obtained from each linear model substituting in (7). This error can be mathematically be written as (92). Obviously, the smaller the difference, the more accurate the model. For the IEEE 30-bus system, the TL-LAC-OPF approach error is 1.8%, which is much less than the error of the LAC-OPF approach (33.9%). In the case of the IEEE 57-bus system, the error of the LAC-OPF approach is 37.7%, while this value for the TL-LAC-OPF approach is only 0.8%. These results confirm that the proposed two-level model significantly reduces the artificial losses caused by the linearization of AC-OPF in SCUC problem.

$$Error = \left| \frac{g_k \theta_k^2 - g_k \sum_{l=1}^{L} k(l) \Delta\theta_k(l)}{g_k \theta_k^2} \right| \times 100 \qquad (92)$$

### 3.1.2. Study B

The system used in this study is the same as the IEEE 30-bus system, except that wind farm and CAES are added to the system. The single line diagram of this system with added wind farm and CAES is shown in Fig. 10. Table VI shows the characteristics of the CAES system located at bus 18. The thermodynamic characteristics of the CAES are considered based on charging and discharging airflow rate given in [22]. In this study, the CAES is considered as a market participant, so that the offered cost for energy, up reserve, and down reserve is equal to 10, 5, and 5 $/MWh, respectively [3]. At the beginning of the day, the existing air in the system is assumed to be 80% of its maximum value. The penalty cost for wind power curtailment, which is imposed on the system operator, is set at 100 $/MWh [3], and the mandatory load shedding cost is 1000 $/MWh. The wind farm is placed on bus 23. Fig. 11 shows the output power of the wind farm located at bus 23 in different scenarios. Moreover, the probability of 15 reduced scenarios from 2000 Monte Carlo scenarios for the output power of the wind farm is shown in Table VII.

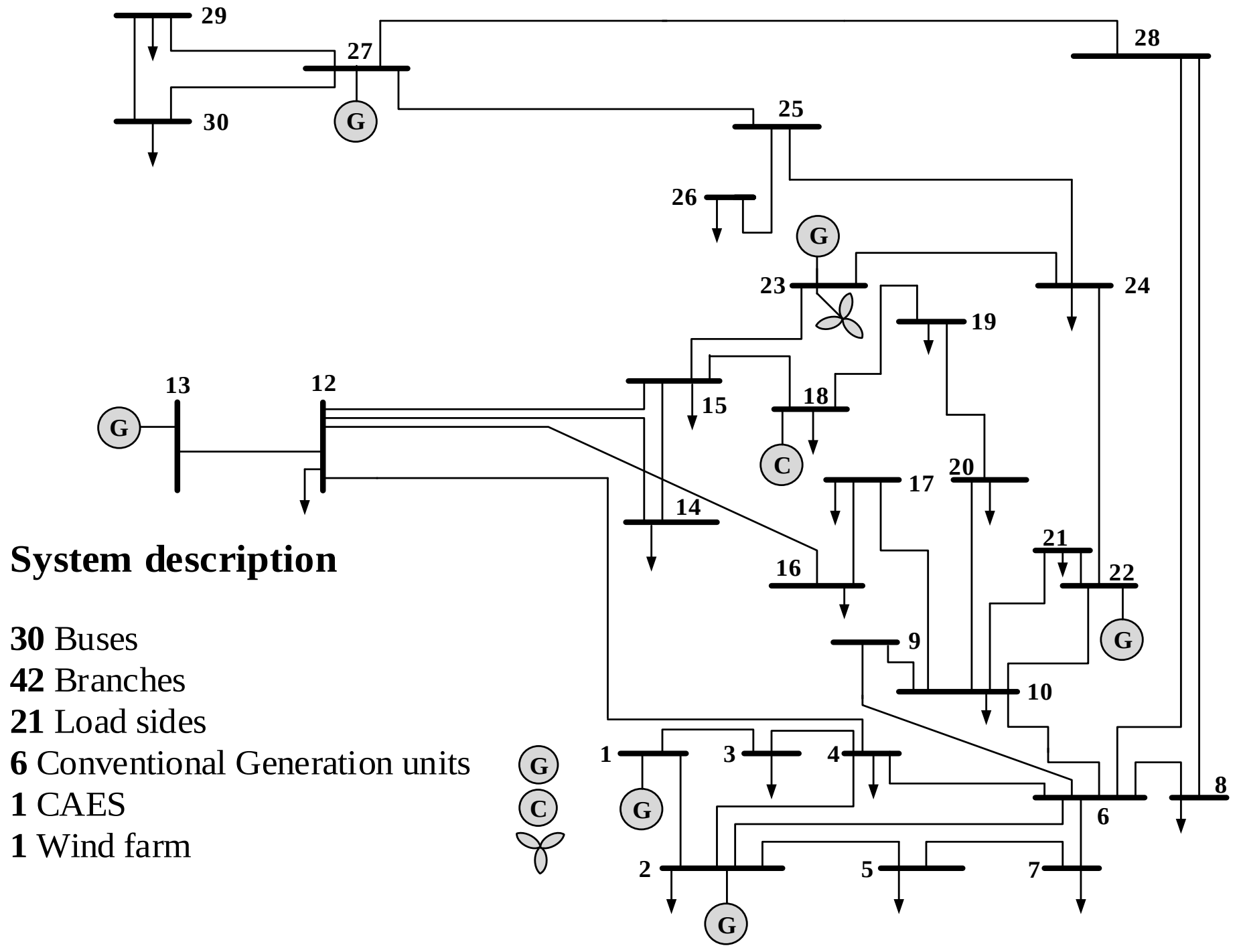

Fig. 10. Single line diagram of the IEEE 30-bus system in the presence of wind farm and CAES.

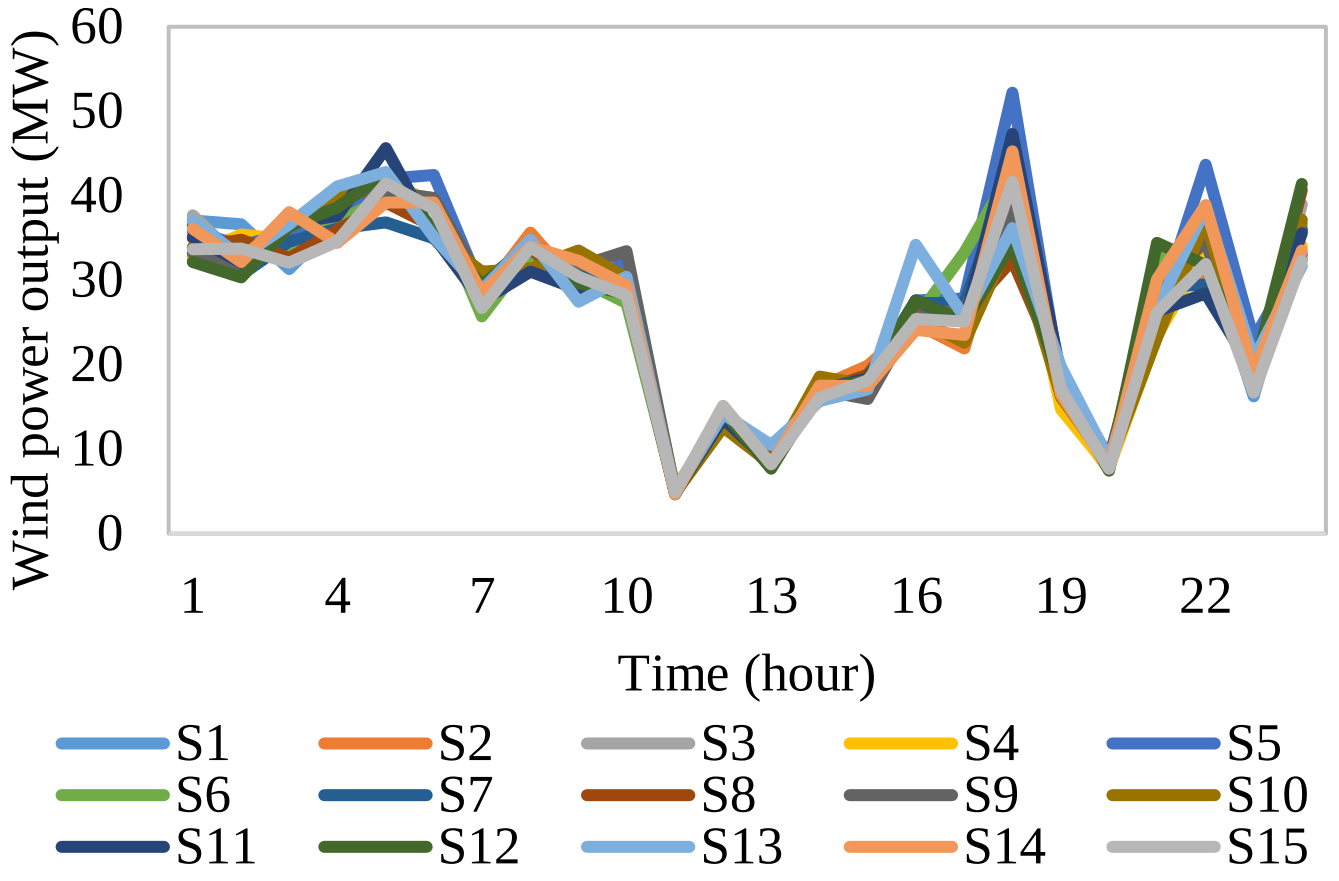

Fig. 11. Output power of wind farm in different scenarios.

Table VI. The characteristics of the CAES.

| $P^{DisCAES,min}$ (MW/h) | $P^{DisCAES,max}$ (MW/h) | $P^{ChCAES,min}$ (MW/h) | $P^{ChCAES,max}$ (MW/h) | $(\%)A^{\min}$ | $(\%)A^{max}$ |
|---|---|---|---|---|---|
| 12 | 40 | 12 | 40 | 33 | 100 |

Table VII. Probability of different scenarios for the output power of wind farm

| Wind scenario # | Probability | Wind scenario # | Probability | Wind scenario # | Probability |
|---|---|---|---|---|---|
| 1 | 0.0673 | 6 | 0.0289 | 11 | 0.0583 |
| 2 | 0.071 | 7 | 0.065 | 12 | 0.0476 |
| 3 | 0.1704 | 8 | 0.05 | 13 | 0.0677 |
| 4 | 0.0749 | 9 | 0.0693 | 14 | 0.0472 |
| 5 | 0.0166 | 10 | 0.095 | 15 | 0.0718 |

#### 3.1.2.1. Comparison of thermodynamic-based model and general model models of CAES in the co-optimized energy and reserve market

In this study, three cases are designed to demonstrate the performance of the proposed model and the effects of thermodynamic-based model (TBM) and general model (GM) of CAES on the co-optimized energy and reserve market:

- Case I: Co-optimized energy and reserve market by considering the wind farm, and without CAES
- Case II: Co-optimized energy and reserve market by considering the wind farm and CAES with GM model.
- Case III: Co-optimized energy and reserve market by considering the wind farm and CAES with the TBM model.

Table VIII lists the total operation costs of these three cases. According to this table, the operation costs without CAES (i.e., \$109972.3) are higher than those of other cases. The presence of the CAES with the GM model reduced the operation costs to \$102750.9. Compared to Case I, there is a decrease of 6.52%, which confirms the cost-effectiveness of this storage device in such markets with AC network constraints. Regarding TBM-based CAES, the operation cost is \$103267.8, approximately 6.09% lower than the that of Case I. It can be inferred from the results that considering the CAES with TBM model increases the costs, compared to the GM model, which happens because of considering more realistic operational conditions of CAES. It can be seen that with the TBM model, this storage can still have relatively significant economic benefits for the power system by reducing costs by 6.09 percent. In this table, the energy cost in case I is equal to the energy cost of thermal units, whereas in cases II and III, the energy cost includes costs of both thermal units and CAES. The cost of supplying energy in the presence of CAES, based on both models, is lower than the costs without this flexible resource. However, considering the TBM model for CAES increases the cost of energy supply from $98029.5 (in Case II) to $98482.5 (in Case III), i.e., $453.025 increase. The reason for the increase in energy costs is due to the more accurate model of CAES which operates in conditions with different efficiency. According to Table VIII, the start-up cost is $1180 for cases II and case III; however, the start-up cost of case I is $1380, which is higher than cases II and case III. This is due to the fact that the presence of CAES increases the flexibility of the system against the changes in power of wind farms. Consequently, the number of start-ups of thermal units for compensating the uncertainty and variability of wind generation decreases. According to Table VIII, since the involuntary load shedding imposes a high cost to the system operator, its cost for all cases is zero. The reserve cost of case I is less than cases II and III. However, on the other hand, in this case I, WPS has occurred, which is not desirable for the system operator, which shows the importance of storage in the maximum use of wind energy generation.

Table VIII. Comparison of operation costs obtained from 3 scenarios

| | Case I | Case II | Case III |
|---|---|---|---|
| Total cost ($) | 109972.3 | 102750.9 | 103267.8 |
| Total cost reduction compared to Case I (%) | - | 6.52 | 6.09 |
| Energy supply cost ($) | 104685.3 | 98029.5 | 98482.5 |
| Start-up cost of thermal units ($) | 1380 | 1180 | 1180 |
| Reserve cost ($) | 3325.3 | 3541.3 | 3605.2 |
| WPS cost ($) | 581.6 | 0 | 0 |
| Involuntary load shedding cost ($) | 0 | 0 | 0 |

Fig. 12 compares the compressed air level in the CAES for TBM and GM models. According to the Fig. 12 (a), the changing/discharging pattern of stored air in the scheduling horizon is relatively similar for both models. However, the stored air level for TBM and GM models in hours 2 to 7 and 14 to 24 is significantly different (see Fig. 12 (b)). Based on the Fig. 12 (b), when the amount of stored air obtained from the GM model is higher than the amount obtained from the TBM model, the maximum difference reaches 17%. If the opposite happens, the maximum difference is -11%.

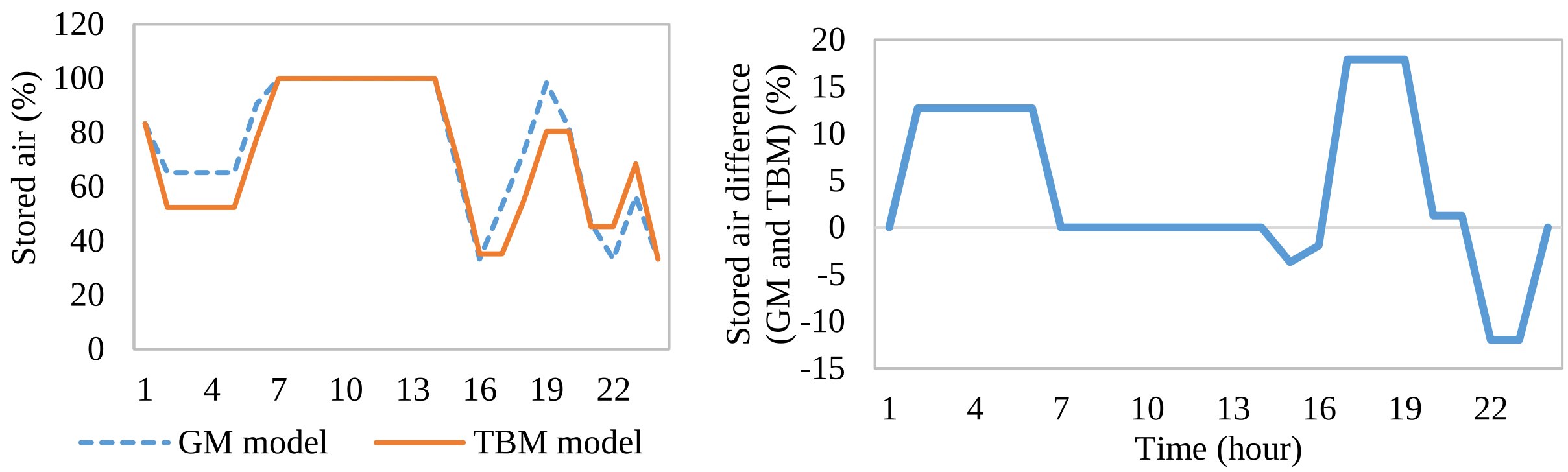

Fig. 12. Comparison of the stored air in CAES with TBM and GM models: (a) actual level comparison, (b) difference comparison.

The effect of the stored air on the charging power of the CAES is shown in Fig. 13. Since the available amount of air in the reservoir directly affects the incoming airflow rate to the CAES, when the storage system tends to be charged in hour 5, the existing pressure in the reservoir (stored air) in TBM model is lower than GM model to prevent reducing efficiency for storing a high charging power. In this way, TMB-based model can store more amount of air by consuming an equal amount of power from the grid. For the TBM model in hour 6, the stored air level is lower because the previous hour required less pressure in the reservoir. On the other hand, the CAES tends to have the maximum capacity for use in the peak hours. Thus, the TMB-based CAES is charged with more power to compensate for the lack of stored air inside the reservoir. In hour 16, the CAES is charged only in the GM model, and not in the TBM model. The reason is that in the next hour, at 17, it is more economical for the system operator to charge the system because the load is lower than that of hour 16. If the CAES is charged at 16 for the TBM model, the reservoir pressure increases, which means that in hour 17 more power with less efficiency is required to store a certain amount of air in the reservoir, which is not economical. Fig. 14 illustrates the output power of CAES obtained from the two models. The comparison of discharged power of the two models shows that small amounts of discharging power, in the TBM model, have been eliminated in some hours, such as 19 and 21. Furthermore, the TBM-based CAES generates more power during the hours 1, 15, and 23. In fact, because of the reduced turbine efficiency and the higher airflow rate required in the low-power operation, it is not economical for the system operator to discharge the TBM-based CAES with low power.

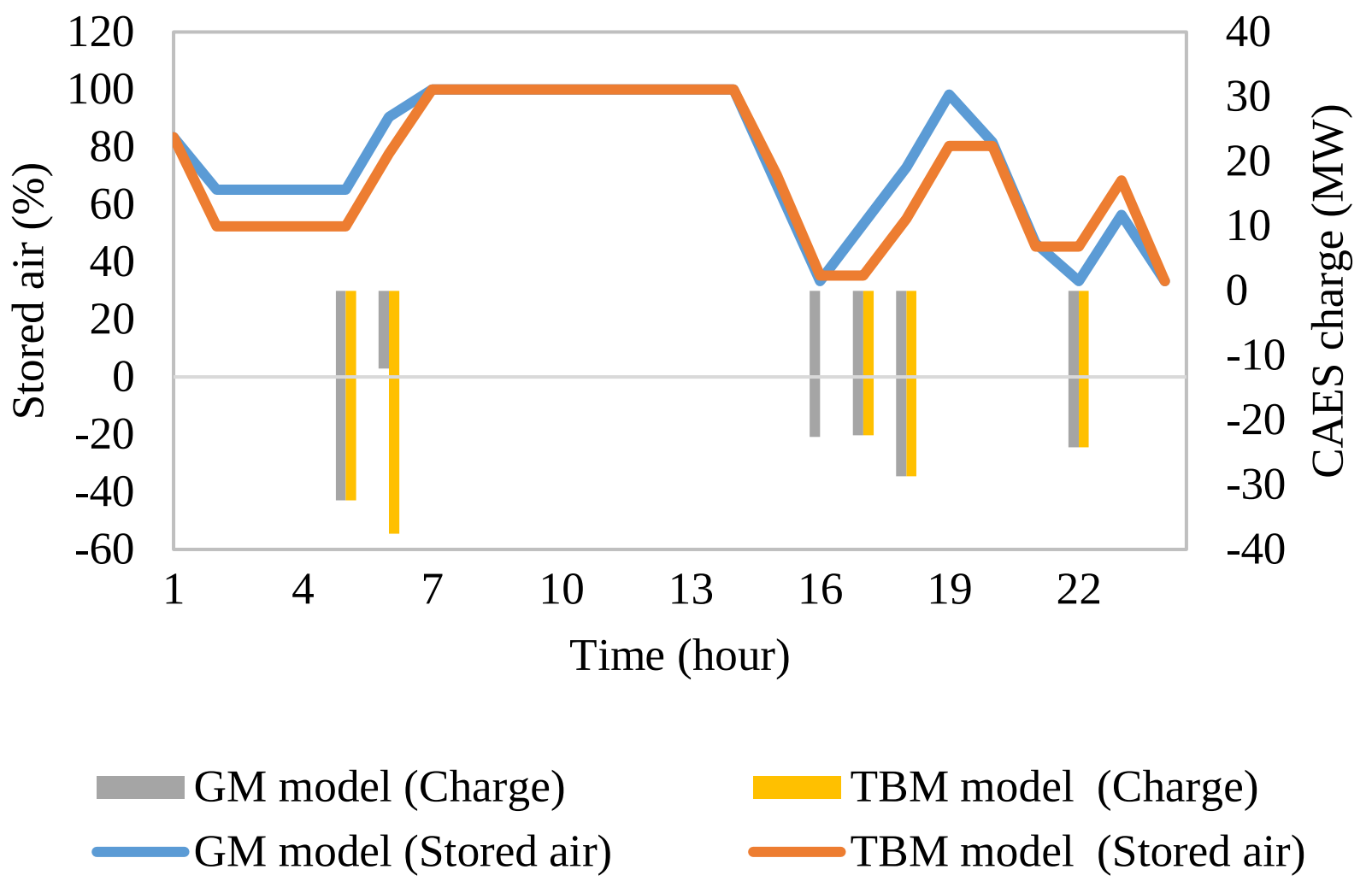

Fig. 13. Effect of the amount of stored air on the charge of CAES.

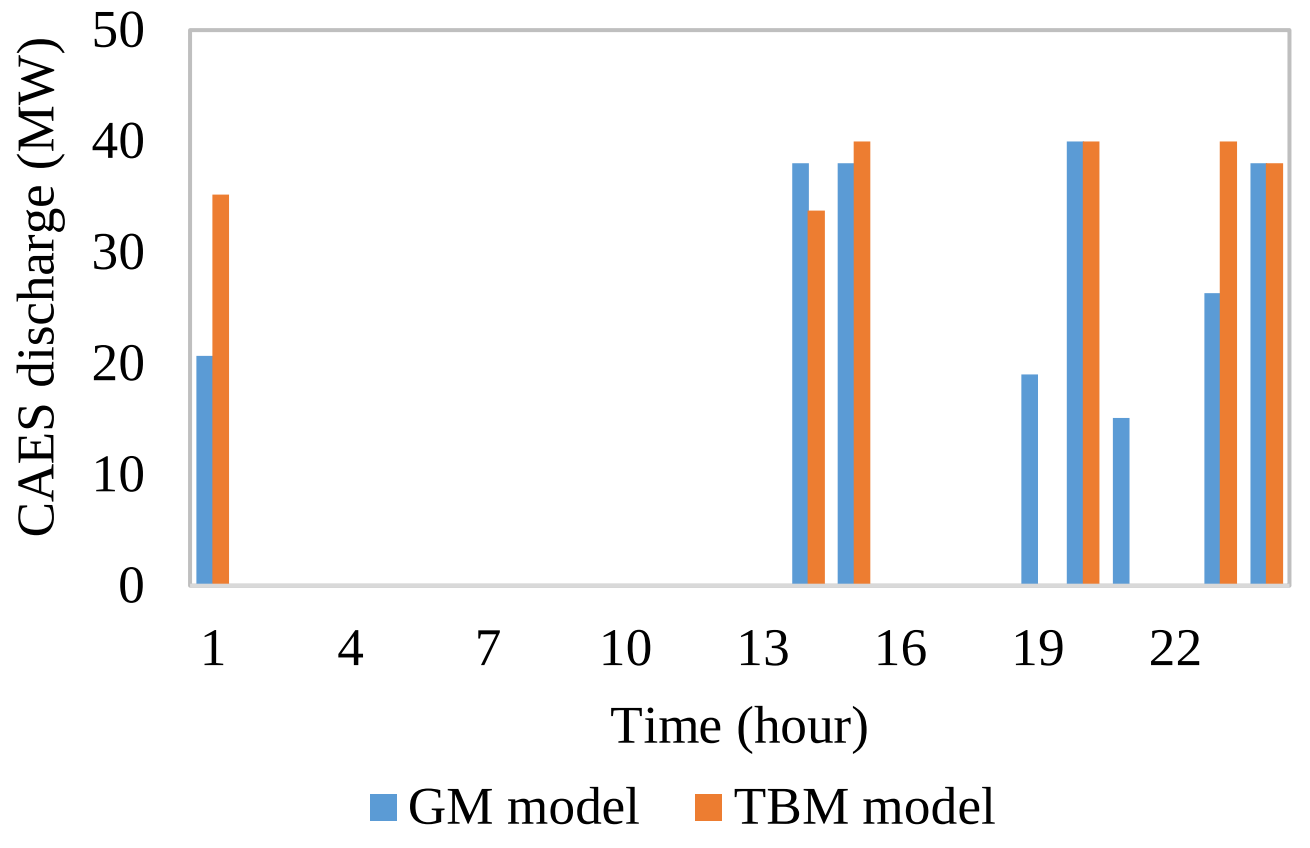

Fig. 14. Comparison of CAES discharge for TBM and GM models.

#### 3.1.2.2. GM model performance with actual thermodynamic condition

In this section, actual operation cost of the GM model is obtained when it comes to implement market decisions with the presence of the CAES thermodynamic conditions. To do so, the TBM model is solved while the commitment of thermal units and charging/discharging of the CAES unit are fixed based on the values obtained from the GM model. This way, one can see that how the dispatch results are changed based on the real condition with the presence of the CAES thermodynamic characteristic. After implementation of this process, the actual operation cost of the GM model is $104196.2, which is $928.3 more than the operation cost of the TBM model, i.e., $103267.8. Thus, although the GM model results in the previous section seem to lead to lower costs, in real conditions, the market results obtained from the TBM model are more economical. In general, the results indicate that considering the thermodynamic characteristics of CAES provides more realistic scheduling, reduces the operation time of CAES, and decreases the network operation costs.

### 3.2. Conclusion

In this paper, a two-stage stochastic DA scheduling framework was proposed to solve linearized AC network-constrained SCUC for clearing a co-optimized energy and reserve market. To reduce the computational time of the proposed optimization problem and increase the accuracy of the LAC-OPF, an engineering insight was applied to convert the proposed model into a two level process, where the first level was an offline process. Then, the proposed two-level two-stage model was developed in order to consider wind farms and large-scale CAES and its thermodynamic characteristics in the co-optimized energy and reserve market, in which the CAES, as a market participant, offers price in the energy and reserve market. The following is a summary of the results of these studies:

- Comparing the two-level and single-level models of linearized AC power flow indicated that the proposed two-level model needs a shorter time to reach to optimal solution, and the artificial losses of the two-level model are much lower than the single-level model. Also, the simulation results proved that by increasing the size of the studied system, the proposed method is able to maintain its performance effectively, which shows the efficiency of the proposed two-level model to solve the AC network-constrained SCUC problem.
- CAES significantly reduces the total operation cost (by at least 6 percent), the commitment of conventional thermal units, the cost of reserve supply, and the start-up cost of thermal units for both GM and TBM models.
- The simulation results of both GM and TBM models of CAES showed that more accurate modeling of this storage is of great importance, because the GM model is inaccurate, and compared to the TBM model, can cause considerable changes in the scheduling of the thermal units, the operation cost of energy and reserve market, charging and discharging levels of CAES, and its compressed air level. Therefore, the market results obtained from the GM model may be different in real conditions.
- Due to the inaccuracy of the GM model of CAES, the scheduling results obtained from this model are not optimal compared to the real conditions that considers the TBM model. Consequently, the real-condition cost of energy and reserve market increases compared to the TBM model.

### References

[1] M. Ghaljehei, Z. Soltani, J. Lin, G. B. Gharehpetian, and M. A. Golkar, “Stochastic multi-objective optimal energy and reactive power dispatch considering cost, loading margin and coordinated reactive power reserve management,” *Electric Power Systems Research*, vol. 166, pp. 163–177, Jan. 2019, doi: 10.1016/j.epsr.2018.10.009.

[2] J. Aghaei, A. Nikoobakht, M. Mardaneh, M. Shafie-khah, and J. P. S. Catalão, “Transmission switching, demand response and energy storage systems in an innovative integrated scheme for managing the uncertainty of wind power generation,” *International Journal of Electrical Power & Energy Systems*, vol. 98, pp. 72–84, Jun. 2018, doi: 10.1016/j.ijepes.2017.11.044.

[3] E. Heydarian-Forushani, M. E. H. Golshan, M. P. Moghaddam, M. Shafie-khah, and J. P. S. Catalão, “Robust scheduling of variable wind generation by coordination of bulk energy storages and demand response,” *Energy Conversion and Management*, vol. 106, pp. 941–950, Dec. 2015, doi: 10.1016/j.enconman.2015.09.074.
[4] M. Ghaljehei, A. Ahmadian, M. A. Golkar, T. Amraee, and A. Elkamel, “Stochastic SCUC considering compressed air energy storage and wind power generation: A techno-economic approach with static voltage stability analysis,” *International Journal of Electrical Power & Energy Systems*, vol. 100, pp. 489–507, Sep. 2018, doi: 10.1016/j.ijepes.2018.02.046.
[5] R. Khalilisenobari and M. Wu, “Optimal Participation of Price-Maker Battery Energy Storage Systems in Energy, Reserve and Pay as Performance Regulation Markets,” in *2019 North American Power Symposium (NAPS)*, Oct. 2019, pp. 1–6, doi: 10.1109/NAPS46351.2019.9000230.
[6] M. Mousavi and M. Wu, “A DSO Framework for Comprehensive Market Participation of DER Aggregators,” in *2020 IEEE Power Energy Society General Meeting (PESGM)*, Aug. 2020, pp. 1–5, doi: 10.1109/PESGM41954.2020.9281462.
[7] A. Nikoobakht, J. Aghaei, and M. Mardaneh, “Optimal transmission switching in the stochastic linearised SCUC for uncertainty management of the wind power generation and equipment failures,” *IET Generation Transmission Distribution*, vol. 11, no. 10, pp. 2664–2676, 2017, doi: 10.1049/iet-gtd.2016.1956.
[8] B. Vatandoust, A. Ahmadian, M. A. Golkar, A. Elkamel, A. Almansoori, and M. Ghaljehei, “Risk-Averse Optimal Bidding of Electric Vehicles and Energy Storage Aggregator in Day-Ahead Frequency Regulation Market,” *IEEE Transactions on Power Systems*, vol. 34, no. 3, pp. 2036–2047, May 2019, doi: 10.1109/TPWRS.2018.2888942.
[9] E. Nasrolahpour and H. Ghasemi, “A stochastic security constrained unit commitment model for reconfigurable networks with high wind power penetration,” *Electric Power Systems Research*, vol. 121, pp. 341–350, Apr. 2015, doi: 10.1016/j.epsr.2014.10.014.
[10] I. Goroohi Sardou, M. E. Khodayar, K. Khaledian, M. Soleimani-damaneh, and M. T. Ameli, “Energy and Reserve Market Clearing With Microgrid Aggregators,” *IEEE Transactions on Smart Grid*, vol. 7, no. 6, pp. 2703–2712, Nov. 2016, doi: 10.1109/TSG.2015.2408114.
[11] B. Cleary, A. Duffy, A. OConnor, M. Conlon, and V. Fthenakis, “Assessing the Economic Benefits of Compressed Air Energy Storage for Mitigating Wind Curtailment,” *IEEE Transactions on Sustainable Energy*, vol. 6, no. 3, pp. 1021–1028, Jul. 2015, doi: 10.1109/TSTE.2014.2376698.
[12] M. Abbaspour, M. Satkin, B. Mohammadi-Ivatloo, F. Hoseinzadeh Lotfi, and Y. Noorollahi, “Optimal operation scheduling of wind power integrated with compressed air energy storage (CAES),” *Renewable Energy*, vol. 51, pp. 53–59, Mar. 2013, doi: 10.1016/j.renene.2012.09.007.
[13] A. Arabali, M. Ghofrani, and M. Etezadi-Amoli, “Cost analysis of a power system using probabilistic optimal power flow with energy storage integration and wind generation,” *International Journal of Electrical Power & Energy Systems*, vol. 53, pp. 832–841, Dec. 2013, doi: 10.1016/j.ijepes.2013.05.053.
[14] A. N. Ghalelou, A. P. Fakhri, S. Nojavan, M. Majidi, and H. Hatami, “A stochastic self-scheduling program for compressed air energy storage (CAES) of renewable energy sources (RESs) based on a demand response mechanism,” *Energy Conversion and Management*, vol. 120, pp. 388–396, Jul. 2016, doi: 10.1016/j.enconman.2016.04.082.
[15] R. Hemmati, H. Saboori, and S. Saboori, “Assessing wind uncertainty impact on short term operation scheduling of coordinated energy storage systems and thermal units,” *Renewable Energy*, vol. 95, pp. 74–84, Sep. 2016, doi: 10.1016/j.renene.2016.03.054.
[16] Z. Soltani, M. Ghaljehei, G. B. Gharehpetian, and H. A. Aalami, “Integration of smart grid technologies in stochastic multi-objective unit commitment: An economic emission analysis,” *International Journal of Electrical Power & Energy Systems*, vol. 100, pp. 565–590, Sep. 2018, doi: 10.1016/j.ijepes.2018.02.028.
[17] H. Daneshi and A. K. Srivastava, “Security-constrained unit commitment with wind generation and compressed air energy storage,” *IET Generation Transmission Distribution*, vol. 6, no. 2, pp. 167–175, Feb. 2012, doi: 10.1049/iet-gtd.2010.0763.
[18] E. Drury, P. Denholm, and R. Sioshansi, “The value of compressed air energy storage in energy and reserve markets,” *Energy*, vol. 36, no. 8, pp. 4959–4973, Aug. 2011, doi: 10.1016/j.energy.2011.05.041.
[19] D. Pozo, J. Contreras, and E. E. Sauma, “Unit Commitment With Ideal and Generic Energy Storage Units,” *IEEE Transactions on Power Systems*, vol. 29, no. 6, pp. 2974–2984, Nov. 2014, doi: 10.1109/TPWRS.2014.2313513.

[20] T. Das, V. Krishnan, and J. D. McCalley, "Assessing the benefits and economics of bulk energy storage technologies in the power grid," *Applied Energy*, vol. 139, pp. 104–118, Feb. 2015, doi: 10.1016/j.apenergy.2014.11.017.
[21] N. Li and K. W. Hedman, "Economic Assessment of Energy Storage in Systems With High Levels of Renewable Resources," *IEEE Transactions on Sustainable Energy*, vol. 6, no. 3, pp. 1103–1111, Jul. 2015, doi: 10.1109/TSTE.2014.2329881.
[22] S. Shafiee, H. Zareipour, and A. M. Knight, "Considering Thermodynamic Characteristics of a CAES Facility in Self-Scheduling in Energy and Reserve Markets," *IEEE Transactions on Smart Grid*, vol. 9, no. 4, pp. 3476–3485, Jul. 2018, doi: 10.1109/TSG.2016.2633280.
[23] H. Falsafi, A. Zakariazadeh, and S. Jadid, "The role of demand response in single and multi-objective wind-thermal generation scheduling: A stochastic programming," *Energy*, vol. 64, pp. 853–867, Jan. 2014, doi: 10.1016/j.energy.2013.10.034.
[24] M. M. R. Sahebi and S. H. Hosseini, "Stochastic security constrained unit commitment incorporating demand side reserve," *International Journal of Electrical Power & Energy Systems*, vol. 56, pp. 175–184, Mar. 2014, doi: 10.1016/j.ijepes.2013.11.017.
[25] E. Heydarian-Forushani, M. E. H. Golshan, and M. Shafie-khah, "Flexible security-constrained scheduling of wind power enabling time of use pricing scheme," *Energy*, vol. 90, pp. 1887–1900, Oct. 2015, doi: 10.1016/j.energy.2015.07.014.
[26] R. Hemmati, H. Saboori, and S. Saboori, "Stochastic risk-averse coordinated scheduling of grid integrated energy storage units in transmission constrained wind-thermal systems within a conditional value-at-risk framework," *Energy*, vol. 113, pp. 762–775, Oct. 2016, doi: 10.1016/j.energy.2016.07.089.
[27] W. Pickard and D. Abbott, "Addressing the Intermittency Challenge: Massive Energy Storage in a Sustainable Future [Scanning the Issue]," *Proceedings of the IEEE*, vol. 100, pp. 317–321, Feb. 2012, doi: 10.1109/JPROC.2011.2174892.
[28] R. Kaviani, M. Rashidinejad, and A. Abdollahi, "A MILP IGDT-based self-scheduling model for participating in electricity markets," in *2016 24th Iranian Conference on Electrical Engineering (ICEE)*, May 2016, pp. 152–157, doi: 10.1109/IranianCEE.2016.7585508.
[29] H. Zhang, G. T. Heydt, V. Vittal, and J. Quintero, "An Improved Network Model for Transmission Expansion Planning Considering Reactive Power and Network Losses," *IEEE Transactions on Power Systems*, vol. 28, no. 3, pp. 3471–3479, Aug. 2013, doi: 10.1109/TPWRS.2013.2250318.
[30] A. Nikoobakht, M. Mardaneh, J. Aghaei, V. Guerrero-Mestre, and J. Contreras, "Flexible power system operation accommodating uncertain wind power generation using transmission topology control: an improved linearised AC SCUC model," *IET Generation Transmission Distribution*, vol. 11, no. 1, pp. 142–153, 2017, doi: 10.1049/iet-gtd.2016.0704.
[31] S. S. Succar *et al.*, "Compressed Air Energy Storage : Theory , Resources , And Applications For Wind Power 8," 2008.
[32] P. Zhao, L. Gao, J. Wang, and Y. Dai, "Energy efficiency analysis and off-design analysis of two different discharge modes for compressed air energy storage system using axial turbines," *Renewable Energy*, vol. 85, pp. 1164–1177, Jan. 2016, doi: 10.1016/j.renene.2015.07.095.
[33] B. Alizadeh, S. Dehghan, N. Amjady, S. Jadid, and A. Kazemi, "Robust transmission system expansion considering planning uncertainties," *IET Generation Transmission Distribution*, vol. 7, no. 11, pp. 1318–1331, Nov. 2013, doi: 10.1049/iet-gtd.2012.0137.
[34] B. Azimian, R. F. Fijani, E. Ghotbi, and X. Wang, "Stackelberg Game Approach on Modeling of Supply Demand Behavior Considering BEV Uncertainty," in *2018 IEEE International Conference on Probabilistic Methods Applied to Power Systems (PMAPS)*, Jun. 2018, pp. 1–6, doi: 10.1109/PMAPS.2018.8440398.
[35] E. Heydarian-Forushani, M. P. Moghaddam, M. K. Sheikh-El-Eslami, M. Shafie-khah, and J. P. S. Catalão, "A stochastic framework for the grid integration of wind power using flexible load approach," *Energy Conversion and Management*, vol. 88, pp. 985–998, Dec. 2014, doi: 10.1016/j.enconman.2014.09.048.
[36] "The GAMS Software Website." http://www. gams.com/dd/docs/solvers/cplex.pdf.
[37] "GAMS Solver Manuals (CPLEX)." http://www.gams.com/.
[38] "IEEE 57-bus system data." https://www2.ee.washington.edu/research/pstca/pf57/ieee57cdf.txt.